\documentclass[a4paper,fleqn,usenatbib]{mnras}

\usepackage{newtxtext,newtxmath}

\usepackage[T1]{fontenc}
\usepackage{ae,aecompl}

\usepackage{graphicx}	
\usepackage{amsmath}	
\usepackage{amssymb}	
\usepackage{multicol}        
\usepackage{bm}		
\usepackage{pdflscape}	
\usepackage{xcolor}

\newcommand{\kms}{\,km\,s$^{-1}$} 

\title[The curious case of II Lup]{The curious case of II Lup: a complex morphology revealed with SAM/NACO and ALMA}

\author[F. Lykou et al.]{Foteini Lykou,$^{1,2,3}$\thanks{Corresponding author: lykoufc@hku.hk}
A.A. Zijlstra,$^{2,4}$\thanks{Hung Hing Ying Distinguished Visiting Professor}
J. Kluska,$^{5,6}$
E. Lagadec,$^{7}$
P.G. Tuthill,$^{8}$\newauthor
A. Avison,$^{4}$
B.R.M. Norris,$^{8}$
Q.A. Parker$^{1,2}$
\\
$^{1}$Department of Physics, The University of Hong Kong, Chong Yuet Ming Physics Building, Pokfulam Road, Hong Kong\\
$^{2}$Laboratory for Space Research, The University of Hong Kong, Block A, Cyberport 4, 100 Cyberport Rd, Cyberport, Hong Kong\\
$^{3}$Institute for Astrophysics, University of Vienna,T\"urkenschanzstrasse 17, A-1180 Vienna, Austria\\
$^{4}$Jodrell Bank Centre for Astrophysics, The University of Manchester, Oxford Road, M13 9PL, Manchester, UK \\
$^{5}$University of Exeter, School of Physics, Stocker Road, Exeter, EX4 4QL, UK  \\
$^{6}$Institute for Astronomy, KU Leuven, Celestijnenlaan 200D B2401, 3001 Leuven, Belgium \\
$^{7}$ Laboratoire Lagrange (UMR 7293), CNRS, Observatoire de la Cote d'Azur, Bd. de l'Observatoire, 06304 Nice Cedex 4, France \\
$^{8}$Sydney Institute for Astronomy, School of Physics, The University of Sydney, NSW 2006, Australia
}

\date{Accepted XXX. Received YYY; in original form ZZZ}

\pubyear{2018}

\begin{document}
\label{firstpage}
\pagerange{\pageref{firstpage}--\pageref{lastpage}}
\maketitle

\begin{abstract}
We present the first-ever images of the circumstellar environment of the carbon-rich AGB star II~Lup in the infrared and sub-mm wavelengths, and the discovery of the envelope's non-spherical morphology with the use of high-angular resolution imaging techniques with the sparse aperture masking mode on NACO/VLT (that enables diffraction limited resolution from a single telescope) and with ALMA. We have successfully recovered images in $Ks$ (2.18\micron), $L'$ (3.80\micron) and $M'$ (4.78\micron), that revealed the non-spherical morphology of the circumstellar envelope around II~Lup. The stellar surface of the AGB star is unresolved (i.e. $\leq30$ mas in $Ks$) however the detected structure extends up to 110 mas from the star in all filters. Clumps have been found in the $Ks$ maps, while at lower emission levels a hook-like structure appears to extend counter-clockwise from the south. At larger spatial scales, the circumstellar envelope extends up to approximately 23 arcsec, while its shape suggests a spiral at four different molecules, namely CO, SiO, CS and HC$_3$N, with an average arm spacing of 1.7 arcsec which would imply an orbital period of 128 years for a distance of 590pc.
\end{abstract}




\section{Introduction}

In the last decade significant steps have been made in understanding stellar mass loss during the asymptotic giant branch (AGB) stage of low- to intermediate-mass stars ($M_\text{init}=0.8-8$~M$_\text{\sun}$). Although the precise mechanisms for mass loss remain contentious,  some interplay between pulsation, dust formation and radiative driving is highly likely. This has been demonstrated, for carbon stars, by dynamic models of \citet{hoefner2003}, \citet{nowotny2005,nowotny2011}, \citet{mattsson2010}, \citet{liljegren2016}, and their applications to high-angular interferometric observations in, e.g., \citet{wittkowski2017}, \citet{rau2017,rau2015}, \citet{ohnaka2007,ohnaka2015}. For oxygen-rich stars, the interplay has been demonstrated by \citet{bladt2015}, and the interferometric observations by \citet{wittkowski2018}, \citet{karovicova2013} and \citet{ohnaka2012}.

For the case of carbon-rich stars in particular ($M\leq4$~M$_\text{\sun}$), it has been suggested that these stars produce high mass-loss yields, and are therefore important players in the chemical enrichment of the interstellar medium in the solar neighbourhood \citep[e.g.,][]{mattsson2010b,schroeder2001}. However, such predictions are also dependent on the metallicity and the model in use \citep[e.g.,][]{karakas2016}. ~\citet[][and references therein]{nowotny2013} have shown that increasing mass-loss rates in carbon-rich AGB stars will result in higher circumstellar reddening, and therefore these stars will appear more obscured \citep[see also][]{liljegren2016} .

On the subject of what shapes the stellar ejecta, the two most-favoured mechanisms are magnetic fields and binarity, although of course these two are not mutually exclusive and hybrid models may also be favoured. If the morphology is governed by binarity, the oxygen-rich and carbon-rich stars should show the same range of morphologies in their circumstellar structures, but confirming this will require a large sample of sources.

The shape of the stellar ejecta beyond the AGB phase, as witnessed in the majority of imaging surveys \citep[e.g.,][]{castro-carrizo2010,lagadec2011}, often departs from spherical symmetry (hereafter, asymmetry). This is more evident in later evolutionary stages, such as the planetary nebula phase where only about 20\% of planetary nebulae have been found to be spherically symmetric \citep{parker2006}. With the advancement of observational techniques, it is now possible to detect such morphological changes even within two stellar radii from the surface of AGB stars.

As part of an on-going effort in detecting asymmetries in the ejecta of AGB stars \citep[e.g.,][]{lykou2015,lagadec2011,tuthill2000,tuthill2000b}, we present here our investigation of one such star, II~Lup (Section~\ref{thestar}). This work is based on single-dish and on interferometric observations (Section~\ref{obs}) and the results are compared with historical findings from speckle imaging (Section~\ref{mysizes}). We present the first-ever images of II~Lup in the near-infrared (Section~\ref{imaging}) and the sub-mm (Section~\ref{alma}) wavelengths. We derive the physical parameters of II~Lup in Section~\ref{physics},  discuss our imaging results in Section~\ref{discussion}, and present our final conclusions in Section~\ref{conclusions}.


\begin{table}
\centering
\caption{II~Lup astrometry.}\label{positions}
\begin{tabular}{lllc}\hline
Survey & RA (J2000) & Dec (J2000) & Epoch\\ \hline
USNO-B1.0 & 15 23 05.154&	-51 25 58.65&1986.7 \\
USNO-B1.0 &15 23 05.214&	-51 25 59.19 & 1987.4\\
PPMXL & 15 23 05.091&	-51 25 58.79 &1989.91\\
PPMXL & 15 23 05.106&	-51 25 58.94&1991.43\\
GSC2.2 &15 23 05.077 & -51 25 58.58&1997.288\\
GSC2.3 & 15 23 05.084 &	-51 25 58.82 &1992.56\\
2MASS & 15 23 05.075	&-51 25 58.73 &1999.441 \\
DENIS &15 23 05.072	&-51 25 58.84 &1998.458\\
SHS & 15 23 05.083& -51 25 58.89&1999.505\\
HST$^a$ &15 23 04.999 &-51 25 58.29 &2004.611\\
{\it Gaia}      DR2 & 15 23 05.0744 &	-51 25 58.881 & 2015.5\\
\hline
\multicolumn{4}{l}{$^a$ Position derived from {\sc sextractor} photometry in the {\em Hubble}}\\ 
\multicolumn{4}{l}{Legacy Archive.}\\
\end{tabular}
\end{table}

\begin{figure}
	\centering
	\includegraphics[width=0.4\textwidth]{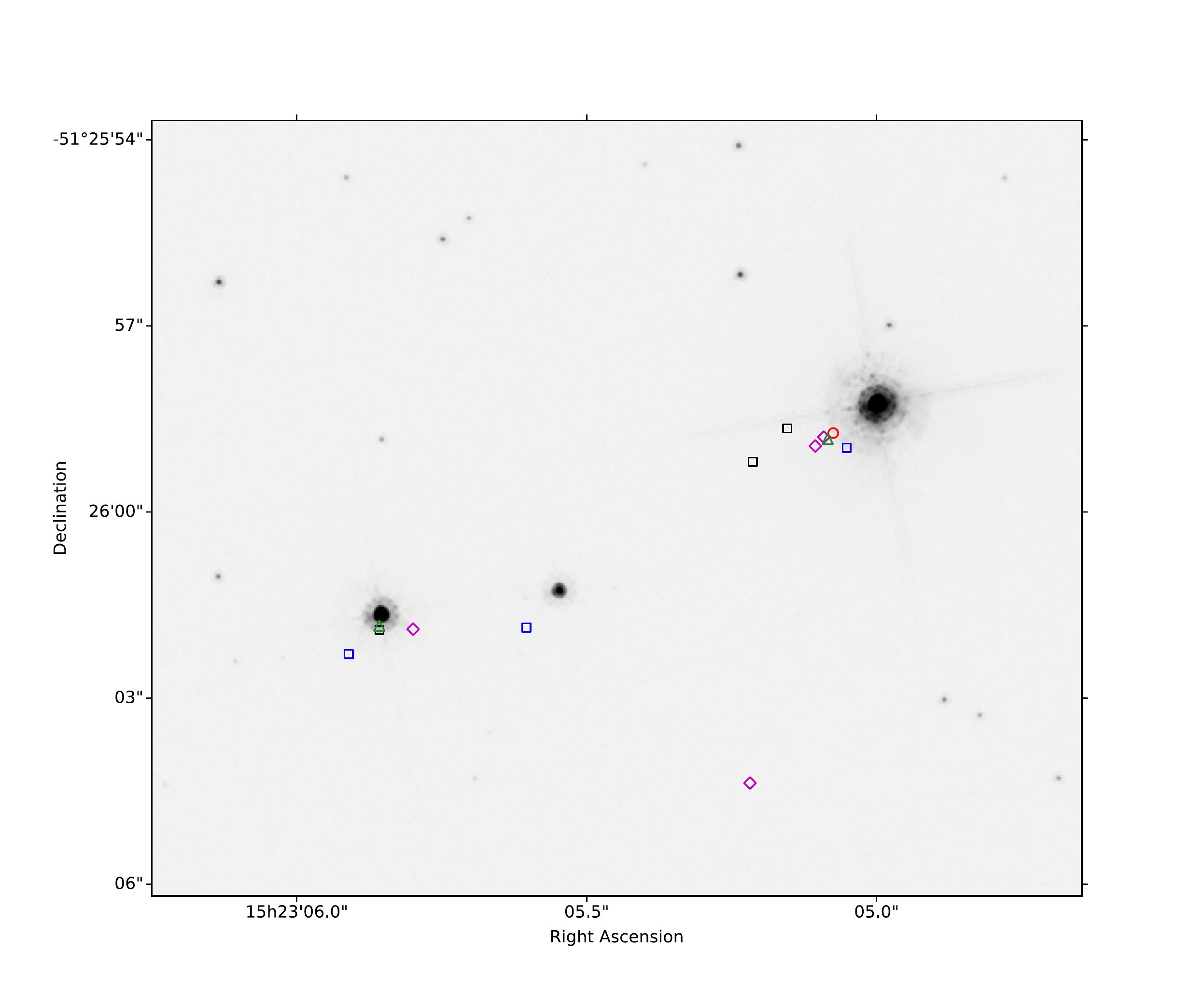}
	\caption{A field image of II~Lup taken with the {\em Hubble} Space Telescope's ACS/HRC camera and the $F606W$ filter. II~Lup is the brightest star in the field, and the region is overlayed with the pointing of several surveys, namely {\it Gaia} DR2 (blue boxes), 2MASS (red circle), GSC2.3 (green triangles), PPMXL (magenta diamonds) and USNO-B.1 (black boxes).}
	\label{hst}
\end{figure}

\begin{figure}
\centering
\includegraphics[width=0.45\textwidth]{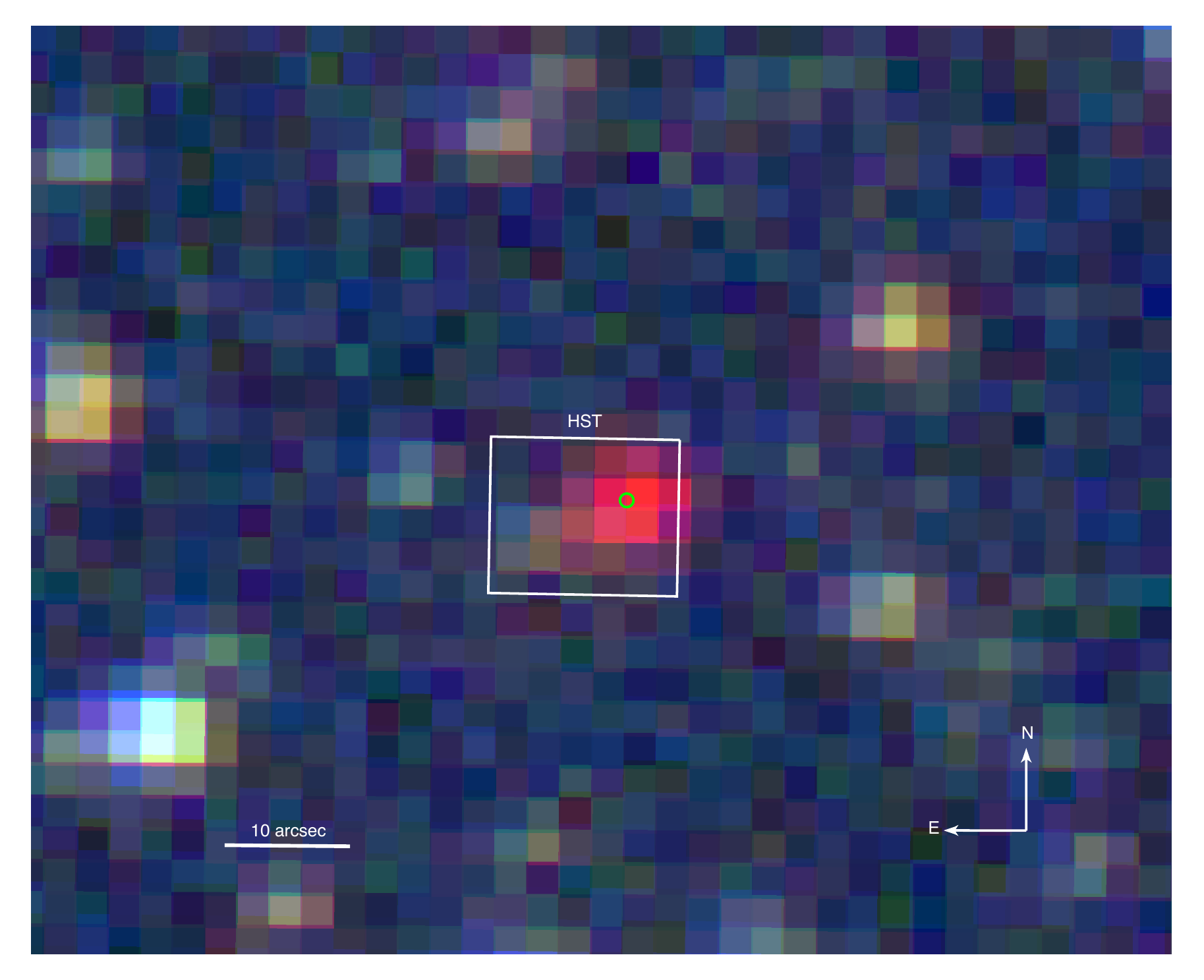}
\caption{A three-colour image of the field around II~Lup from the AAVSO Photometric All-Sky Survey (APASS) in 2011. The colours blue, green and red stand for the filters Johnson $B$, $V$ and Sloan $i'$, respectively. The entire field-of-view shown in Fig.~\ref{hst} is indicated by a white box for comparison. II~Lup was not detected in filters $B$ and $V$ (probably fainter than 16.5 mag in $V$). The oblate morphology of the source within the box is in part due to the poor angular resolution and focus of the APASS survey, and due to the nearby, eastern field stars in Fig.~\ref{hst} which are also bright in $i'$. The position of II~Lup in the {\it Gaia} system is marked with a green circle.}
	\label{apass}
\end{figure}

\section{The case of II~Lup}\label{thestar}
The star II~Lup was known as the third brightest AGB star in the southern hemisphere at 12\micron\ when observed by {\em IRAS} \citep{meadows1987,nyman1993}. It was first mentioned as WO48 in the survey of \citet{westerlund1978} and misclassified as an S-star. It was later observed by the IRAS space telescope \citep[IRAS~15194--5115;][]{iras} and simultaneously identified in the Valinhos infrared survey \citep{epchtein1987}. \citet{meadows1987} characterised II~Lup as a C-rich star based on the carbonaceous species found in its near-infrared spectrum (3.1\micron\ feature).

\subsection{On the astrometry and distance.}\label{distance}
Although a bright infrared source, there is no optical identification published. II~Lup is not included in the Tycho and {\em Hipparcos} catalogues \citep{hipparcos}, but it is included in {\it Gaia} DR2~\citep{gaia,gaia2018}. However, the current {\it Gaia} parallax ($\pi=0.52\pm0.28$ mas) is not reliable, since parallax estimation is problematic for extremely red and variable sources such as II~Lup \citep[][and references therein]{mowlavi2018}. Figure~\ref{hst} shows the field near II~Lup as viewed by the {\em Hubble} Space Telescope (HST) at 0.6$\rm\mu m$ (ACS/HRC filter $F606W$; prog.ID 10185). We extracted astrometric information on II Lup using the {\tt Aladin}\footnote{aladin.u-strasbg.fr/aladin.gml} tool \citep{aladin1}. 

Comparing the coordinates given for the field stars and II~Lup from five major catalogues, namely {\it Gaia} \citep{gaia,gaia2018}, 2MASS \citep{2mass}, PPMXL \citep{ppmxl}, GSC2.3 and USNO-B.1 \citep{usnob}, to the HST image in Figure~\ref{hst} clearly shows the mismatch between the various astrometric solutions (cf. Table~\ref{positions}).  The difference in coordinates of the two bright field stars south-east of II~Lup between the HST and the {\it Gaia} systems are less than 1 arcsec , which is similar to the known variations, depending on the field, of the astrometric uncertainty of the Guide Star Catalogue. The uncertainties both in the coordinates and proper motions of the remaining abovementioned catalogues, indicates that they have all detected the same star: II~Lup. Thus, there is an optical counterpart to this deeply embedded star.


Taking into account the inefficacy of the current {\it Gaia} algorithm in determining accurate parallaxes for variable stars like II~Lup, if we consider the $3\sigma$ limit of the {\it Gaia} DR2 parallax, we do not expect II~Lup to be closer than 500 pc. We have adopted the distance of 590~pc proposed by \citet{groenewegen2002a} and \citet{menzies2006}. This distance was estimated from the period-luminosity relation for carbon-rich Miras of \citet{whitelock2006}.

\begin{table}
\centering
\caption{Photometry of II~Lup.}\label{phot}
\begin{tabular}{llll}\hline
Band & \multicolumn{2}{l}{$m$} & Reference\\ 
$B$ & \multicolumn{2}{l}{17.85} & NOMAD \\
$V$ &\multicolumn{2}{l}{16.66} & NOMAD\\
$R$ & \multicolumn{2}{l}{14.51} & NOMAD\\
$I$ & \multicolumn{2}{l}{10.54} & DENIS \\
$J$ & \multicolumn{2}{l}{7.28}  & (1),(2) \\
$H$ & \multicolumn{2}{l}{4.58}&(1),(2)\\
$K$ & \multicolumn{2}{l}{2.35}&(1),(2)\\
$L$ & \multicolumn{2}{l}{-0.56}&(1)\\
$M$ & \multicolumn{2}{l}{-1.49}&(1)\\ 
$\rm \lambda$~(\micron) & $F$ (Jy) & $F_\text{err}$ (Jy) & Reference\\ 
3.08  & 459 & 12 & {\em ISO} \\
3.35  & 37.5 & 0.3 & {\em WISE} \\
3.35  & 1255.2 & -- & {\em AllWISE} \\
3.5  & 203.3&30.5 & {\em COBE} \\
4.29  & 371.4 &31 & {\em MSX}\\
4.6  & 83.4 & 0.3 & {\em WISE} \\
4.6  & 78.7 & -- & {\em AllWISE} \\
4.9  & 500.6&36.4 & {\em COBE} \\
6.3  & 1000& 100& {\em ISO} (3) \\
8.28  & 193.1 &7.9 & {\em MSX}\\
8.8   & 860 & 86& {\em ISO} (4)\\
9.8  & 852& 85& {\em ISO} (4)\\
11.6  & 427.6& 5& {\em WISE} \\
11.6  & 574.2& --& {\em AllWISE} \\
11.7  & 860& 86& {\em ISO} (4)\\
12  & 1310 &104.8 & {\em IRAS} \\
12  &818.8 &55.5 & {\em COBE} \\
12.13  & 861.8 &43.1 & {\em MSX} \\
12.5  & 681& 68& {\em ISO} (4)\\
14.65  & 630.7 & 38.4& {\em MSX} \\
18  & 526.5 &38.1 & {\em AKARI} \\
21.24  & 488.3& 29.3& {\em MSX} \\
22.1  &676 &-- & {\em WISE} \\
22.1  & 248& 0.45& {\em AllWISE} \\
25  & 553.7&33.2 & {\em IRAS} \\
25  &606.8 &51.5 & {\em COBE} \\
60  &225.7 & 101& {\em COBE} \\
60  & 129.9&12.99 & {\em IRAS} \\ 
65  &107.9 &8.35 & {\em AKARI} \\
70  & 106& 10.6& {\em Herschel} (5)\\
90  & 62.9 & 1.67 & {\em AKARI} \\
100 & 52.57& 6.3& {\em IRAS} \\
140  & 16.14& 1.17& {\em AKARI} \\
160  & 14.97& 1.03 & {\em AKARI} \\
160  &22.9 & 3.4& {\em Herschel} (5)\\
1200  &0.3106&0.019& (6) \\
\hline
\multicolumn{4}{l}{{\bf References:} (1) \citet{lebertre1992}; (2) \citet{whitelock2006};}\\
\multicolumn{4}{l}{(3) \citet{neufeld2011}; (4) \citet{guandalini2006}; }\\
\multicolumn{4}{l}{(5) \citet{groenewegen2011}; (6) \citet{dehaes2007}}\\
\end{tabular}
\end{table}

\begin{table}
\caption{Modelled stellar and envelope radii from the literature converted to angular size for an adopted distance of 590 pc.}\label{envelopes}
\centering
\begin{tabular}{cccc}
\hline
 $R_*$ (au) & $R_\text{env}$ (au) & Reference & $\theta_\text{env}$ (mas) \\ 
\hline
 1.86 & 10.02&(1)&16.99\\
 3.00 & 12.03 & (2) & 20.38\\
 2.54 & 7.19 &(3)&12.19\\
 2.66 & 27.20 & (4)& 46.10\\
 2.52 & 10.02&(5)&16.99\\
 2.32 & 16.48 & (6) &27.93\\
 3.54 & 21.24&(7)&36.00\\
 2.52 & 14.70&(8)&24.92\\
\hline
\multicolumn{4}{l}{{\bf References: } (1) \citet{ramstedt2014};}\\
\multicolumn{4}{l}{(2) \citet{ryde1999}; (3) \citet{debeck2010};}\\
\multicolumn{4}{l}{(4) \citet{dehaes2007}; (5) \citet{schoeier2007}; }\\
\multicolumn{4}{l}{(6) \citet{groenewegen1998}; (7) \citet{lopez1993};}\\
\multicolumn{4}{l}{(8) \citet{danilovich2015}}\\
\end{tabular}
\end{table}

\begin{figure*}
\centering
\includegraphics[width=0.5\textwidth, angle=270,trim=-2cm 0.5cm -1cm 0,clip]{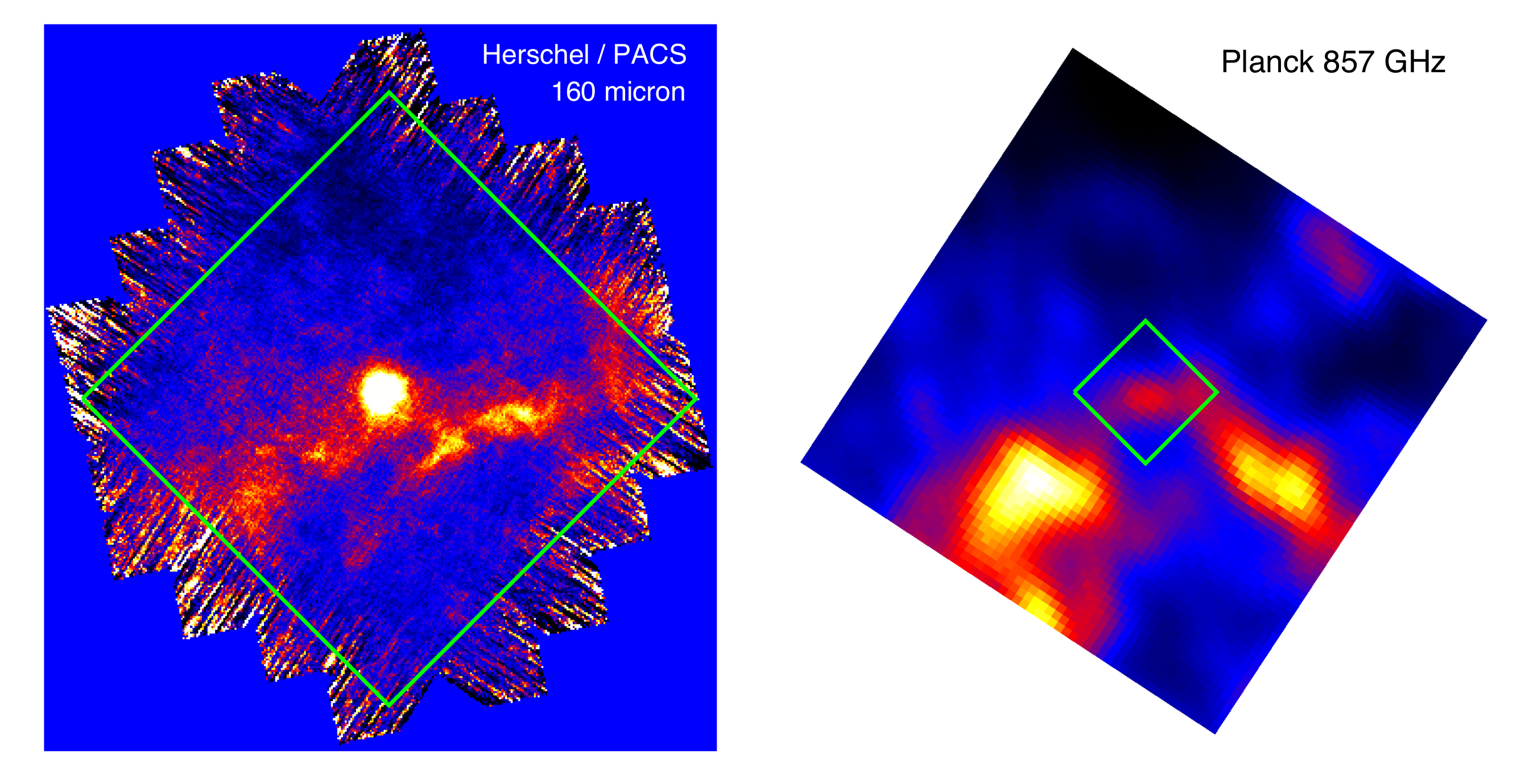}
\caption{Far-infrared region maps of II~Lup from the {\em Herschel} (160\micron ; left) and {\em Planck} (857GHz; right) space telescopes. Some filaments of the background Lupus cloud are clearly visible in the {\em Planck} map (size 1\degr$\times\,$1\degr) with brighter clumps seen more than 0.5\degr\ south-east and south-west from II~Lup (centre). The green boxes are 0.2\degr$\times\,$0.2\degr\ in both panels. North is up and east is left.}\label{farirmaps1}
\end{figure*}

\subsection{Photometry}

There is no available time-series photometry for this star in the visual\footnote{No data found in the archives of AAVSO, ASAS, OGLE or {\em Hipparcos}.}. The only available photometry in $V$ from the literature is found in the work of \citet{meadows1987}, where the authors give an upper limit of $V>21$, and in the NOMAD catalogue \citep[$B=17.85$ and $V=16.66$;][]{nomad}. The AAVSO Photometric Sky Survey (APASS) did not detect the star in either $B$ or $V$, suggesting that the star was fainter than 16.5 mag in $V$ in 2011 (see also Fig.~\ref{apass}). II~Lup is included in the catalog for optically variable sources of the Optical Monitoring Camera (OMC) of the {\em INTEGRAL} space telescope\footnote{sdc.cab.inta-csic.es/omc/index.jsp} \citep{integral}, which estimated $15.5\lesssim V \lesssim14.4$ over a period of several years (2003--2016). However OMC has a pixel size of 17.6 arcsec and it is mentioned that the results were of poor photometric quality. We note that the stellar field of Fig.~\ref{hst} falls within one OMC pixel, which may explain the poor photometric quality\footnote{Bad centroid or centroid too far from the source's coordinates.}.


 The reported colour for II~Lup in the {\it Gaia} system is $G_{\rm BP} - G_{\rm RP}=6.26$ and $G=14.262$. The given magnitudes for II~Lup in the Hubble Legacy Archive is $[F606W]=15.74$ and $[F814W]=12.26$, which is clearly the brightest star in the field in Fig.~\ref{hst}. In Table~\ref{phot}, we have compiled the available photometry from the literature\footnote{Only the mean $R,I,J,H,K,L,M$ photometry is tabulated for the sake of clarity, and only known flux uncertainties are listed.}.

\subsection{On the circumstellar environment in the infrared}\label{environment}

The chemical composition of II~Lup has been extensively studied through terrestrial single-dish observations in the far-infrared and radio regimes. Various radiative transfer models, that assumed a spherical geometry for the circumstellar envelope, have been applied to these observations \citep{nyman1993,ryde1999,groenewegen1999,groenewegen2002b,schoeier2002,dehaes2007,debeck2010,neufeld2011,csmith2012,ramstedt2014}. The results of the modelling vary depending on the adopted distances and luminosities, as well as the radiative transfer model in use. However there is an overall agreement on the {\it gas} mass loss which is $5-20\times10^{-6}$ M$_\text{\sun}$ yr$^{-1}$. 
On the other hand, values for the {\it dust} mass loss vary from $1-8\times10^{-8}$ M$_\text{\sun}$ yr$^{-1}$\citep{groenewegen1998,groenewegen2002b,csmith2012}, again depending on the radiative transfer model applied on the spectral energy
distribution. The modelled stellar radii and inner radii of the envelope ($R_\text{env}$) derived from the literature are tabulated in Table~\ref{envelopes} for reference. We have converted the inner radii $R_{\rm env}$ into angular sizes ($\theta_\text{env}$) for an adopted distance of 590~pc.

We searched for any available image of II~Lup in all (space and ground) telescope data archives in the infrared regime. Due to the star's extreme brightness in the near- and mid-infrared, the images found in the DENIS, 2MASS, {\em WISE} \citep{wright2010}, {\em MSX} and {\em IRAS} archives, are either saturated or indicate that II~Lup was unresolved. In the far-infrared, the circumstellar envelope of II~Lup is resolved in the {\em AKARI} \citep{murakami2007}, {\em Herschel} \citep{herschel} and {\em Planck} \citep{planck} images beyond 65\micron. These images also reveal background emission from the Lupus interstellar cloud (Fig.~\ref{farirmaps1}). II~Lup is much brighter than the background emission in all images. This may be further justified by the lower extinction calculated by \citet{menzies2006} ($0.44\leq A_V\leq 0.48$) and \citet{groenewegen1998} ($A_V=0.8$) compared to that of \citet{schlaflyfink2011} ($A_V=2.38$). The last is based on an analysis of the Sloan Digital Sky Survey, as well as far-infrared temperature maps from the DIRBE/{\em COBE} and {\em IRAS} surveys. The size of II~Lup's circumstellar cloud as determined from the 70\micron\ image is approximately 40 arcsec , however it bears no distinctive morphological features other than a round shape
\citep{cox2012,groenewegen2011}. 

\begin{figure}
\centering
\includegraphics[width=1.1\columnwidth]{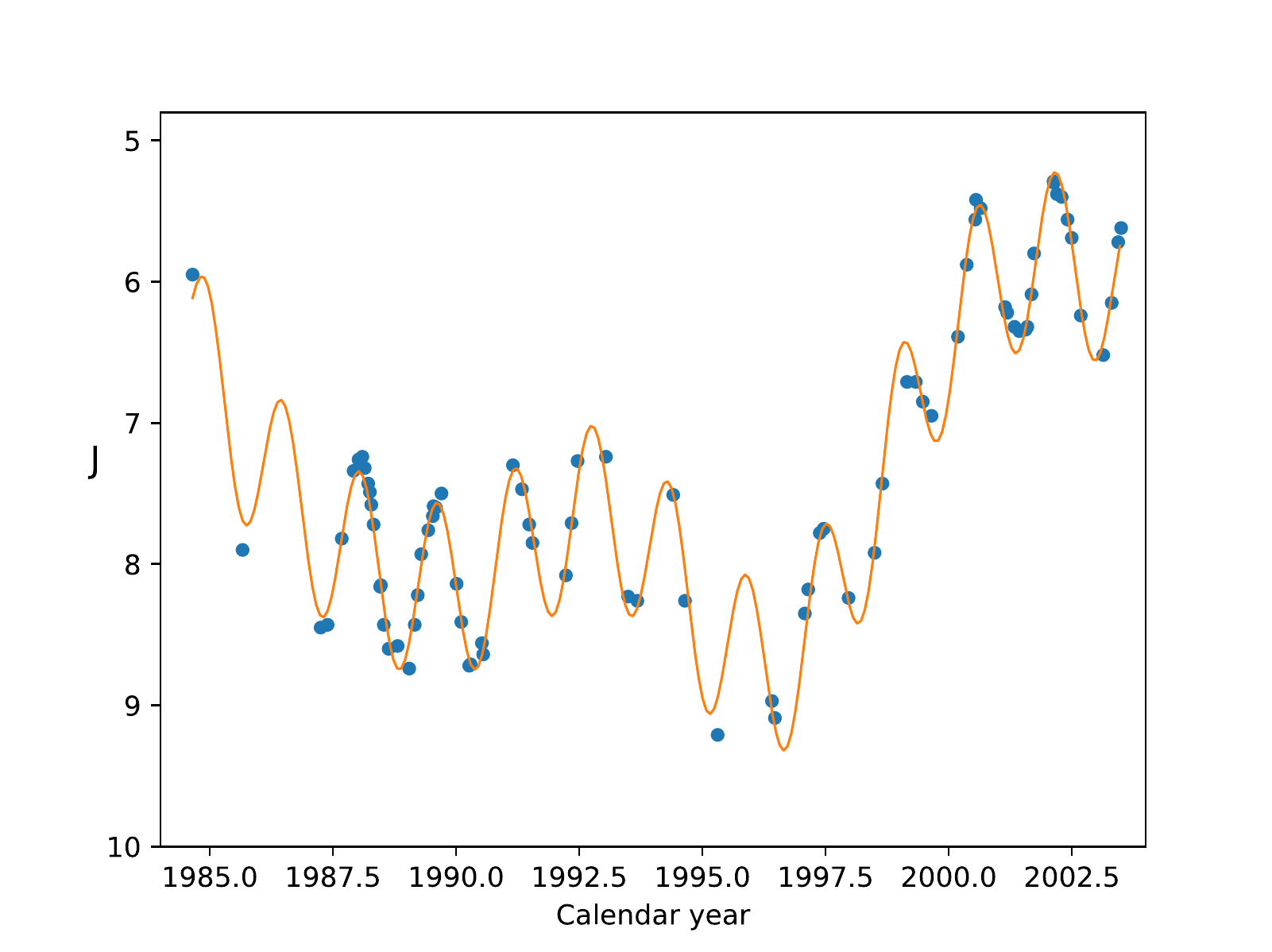}
\caption{{\sc period04} fit to the $J$ band light curve. Photometry extracted from: \citet{epchtein1987,lebertre1992,feast2003,whitelock2006}. }\label{lc}
\end{figure}
 \begin{figure*}
 \centering
\begin{minipage}{.33\textwidth}
        \includegraphics[scale=0.45]{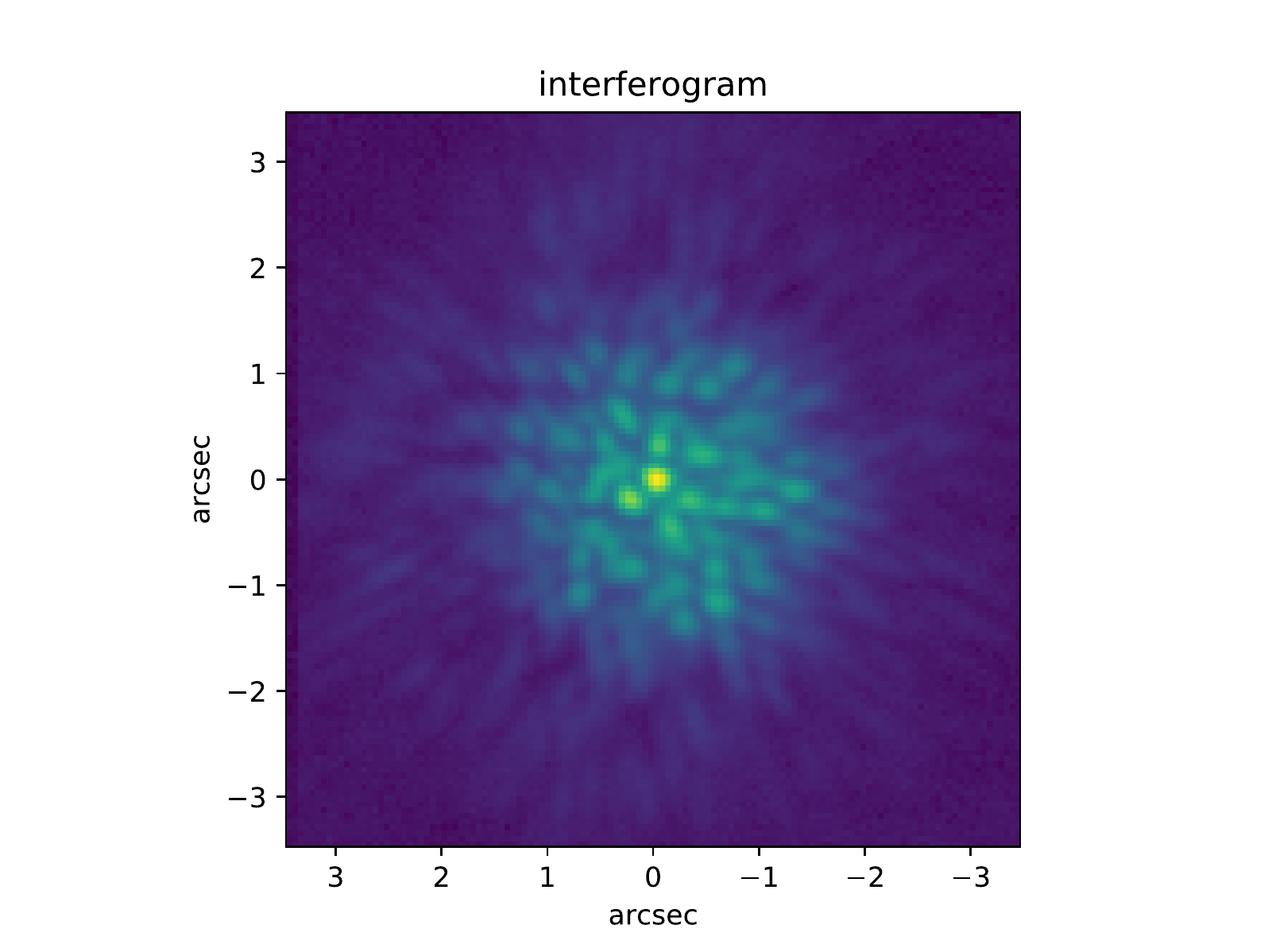}
   \end{minipage}%
    \begin{minipage}{.33\textwidth}
        \includegraphics[scale=0.45]{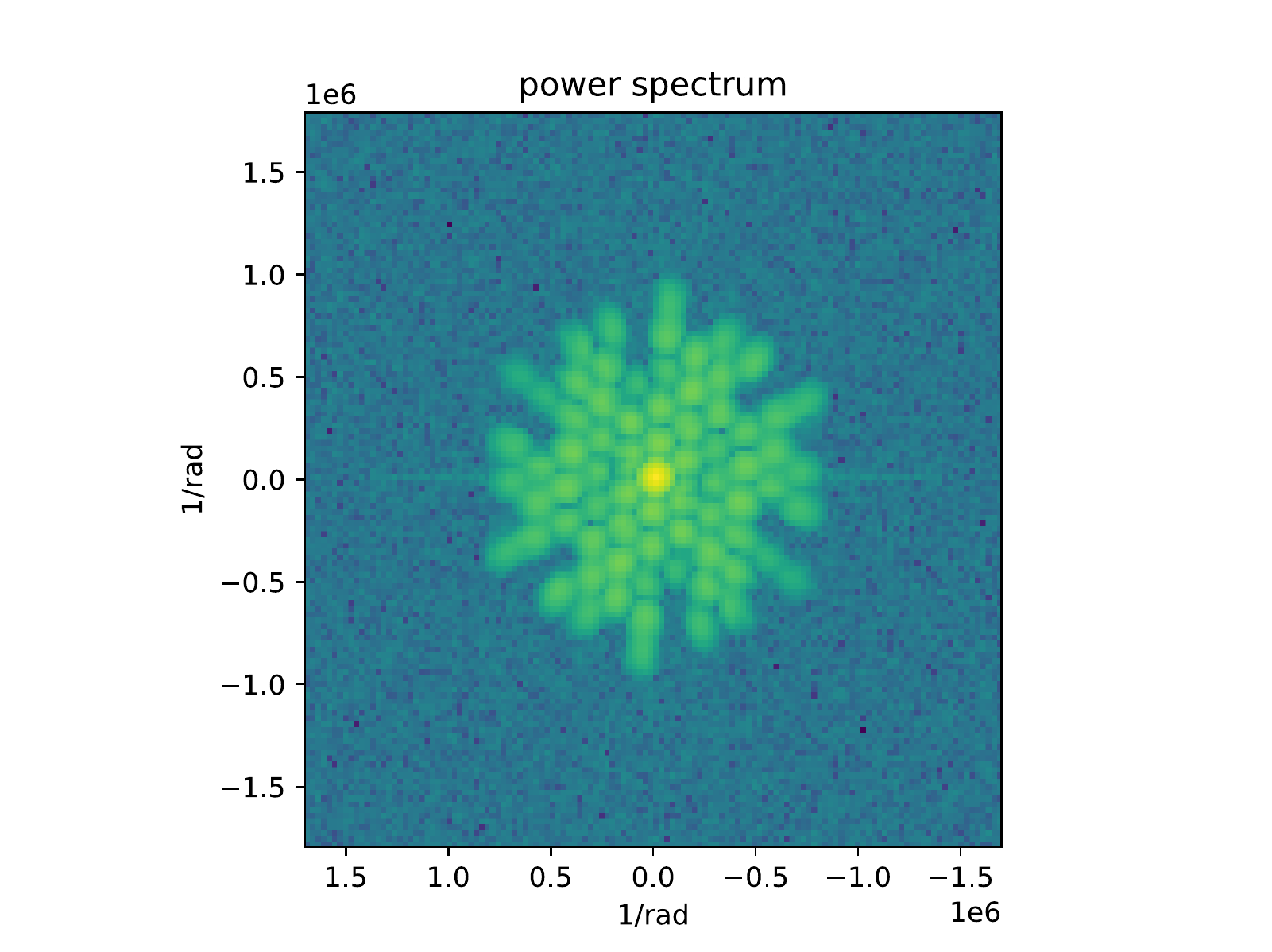}
    \end{minipage}%
    \begin{minipage}{.33\textwidth}
        \includegraphics[scale=0.45]{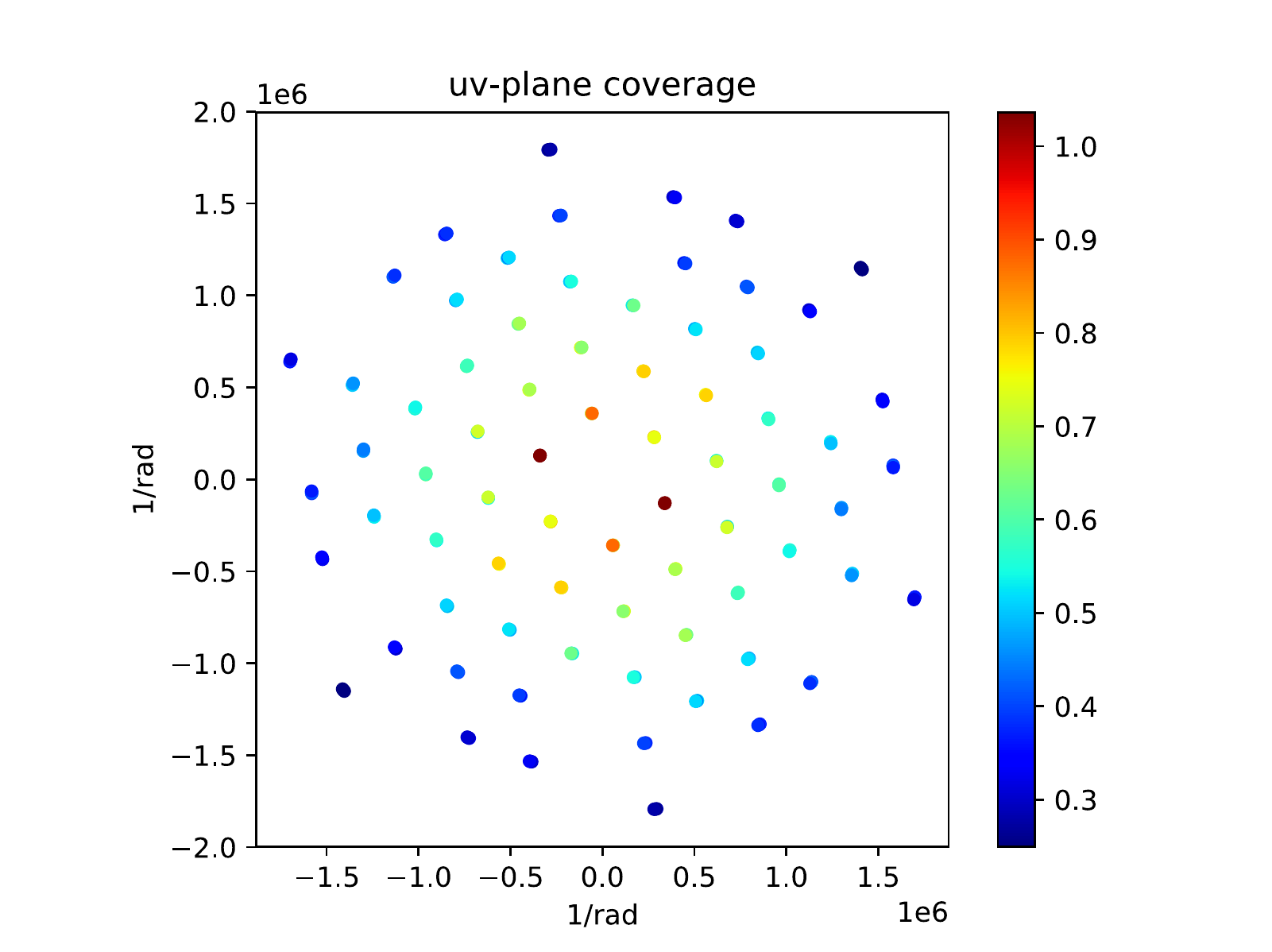}
    \end{minipage}%
\caption{{\it Left:} An example of the interference pattern created by the 36 apertures of the 9-hole mask. The first Airy ring is visible (online version of this figure) suggesting that the object was resolved. North is up and east is on the left. {\it Centre:} the power spectrum of the given interferogram. Each spot corresponds to the sampled visibility and its complex conjugate. {\it Right:} the $uv$-plane coverage with the 9-hole mask for II~Lup. The points are colour-coded with increasing values of squared visibility amplitude.}\label{speckles}
 \end{figure*}

 \subsection{On the peculiar secondary period}\label{period}

II~Lup has been monitored in the near-infrared ($J$, $H$, $K$, $L$ and $M$) over the course of approximately 20
years \citep{epchtein1987,lebertre1992,feast2003,whitelock2006}. \citet{feast2003} presented the first comprehensive study of II~Lup's photometric variability. From their analysis, the authors derived a primary period of 575 days and a secondary period of approximately 19 years (6900 days); with the longer period proposed to result from an {\it obscuration event}. Such events have been observed before in other AGB stars and appear to be relatively common. Prominent examples are the stars V~Hya \citep{knapp1999}, R~Lep, R~For, EV~Eri \citep{whitelock2006}, L2~Pup \citep{bedding2002}, and  M2--29 \citep{hajduk2008}.

\citet{feast2003} argue that the extreme variability at the shorter wavelengths ($J$ to $K$) is due to scattering by dust, which was  suggested by \citet{lopez1993} as well. \citet{feast2003} suggested the presence of dusty puffs around II~Lup. 
From its {\em IRAS} colours ($[12]-[25] = -0.921$, and $[25]-[60]= -1.476$) and its intrinsic variability in the near-infrared, II~Lup is somewhat similar to a class of variable carbon-rich stars known as R~CrB stars. The principle star of this class, R~CrB, has similar {\em IRAS} colours ($-0.924$ and $-1.57$, respectively) and is also known to have ejected large dust puffs in the past \citep{jeffers2012,clayton2011}. However, the variability of II~Lup is not as rapid as that of R~CrB, where typical primary pulsation periods of 40 days are found in the visual light curves \citep{crause2007}.

According to \citet{whitelock2006} obscuration events, such as the one proposed for the case of II~Lup, are found in one-third of carbon-rich Miras in their sample. According to \citet{olivier2003}, such long secondary periods can be explained either by an expulsion of dusty clumps \citep[such as those observed in R CrB stars: ][]{jeffers2012,clayton2011} or by the presence of a companion. \citet{olivier2003} and \citet{wood2009} have confirmed the binary nature of many stars with long secondary periods in the LMC, while at least two cases are known within our Galaxy: e.g. L2 Pup~\citep{kervella2016} and V Hya \citep{knapp1999}.

Additional time-series infrared photometry was found in the DIRBE/{\em COBE} \citep{price2010} and {\em WISE} survey databases. \citet{price2010} suggest a period of 632 days from the $L$ and $M$ DIRBE timeseries. This period is significantly longer than the derivations of \citet{feast2003} and \citet{menzies2006}. We note that {\em WISE} photometry is available for only two epochs and may suffer from saturation in all bands. 

We have re-analysed all available infrared time-series photometry from the literature with {\sc period04} \citep{lenz2004}, a software commonly used in asteroseismology. From this new analysis and the inclusion of the DIRBE/{\em COBE} data, the primary period of 575 days derived by \citet{feast2003} is confirmed (this work: 574$\pm$2 days), as well as the secondary period (this work: 19.7$\pm$0.2 years). Variations were found in the secondary period in $H$ and $K$ bands ($\simeq25$ years), however \citet{feast2003} did not report any secondary period calculations in those bands. {\sc period04} provided a good fit to the $J$ band photometry (Fig.~\ref{lc}), while the distinctive brightening after 1996, as mentioned by \citet[][their fig.1]{feast2003}, is also satisfied by the synthetic light curve of {\sc period04}. A variation in the calculation of the secondary period was expected. This is caused mainly by the uneven sampling in the time series especially from 2003 until now. Below we will argue that the orbital period is much longer;  the derived  secondary period may therefore reflect the limited time coverage rather than the true value. 

\begin{figure}
\includegraphics[width=.45\textwidth]{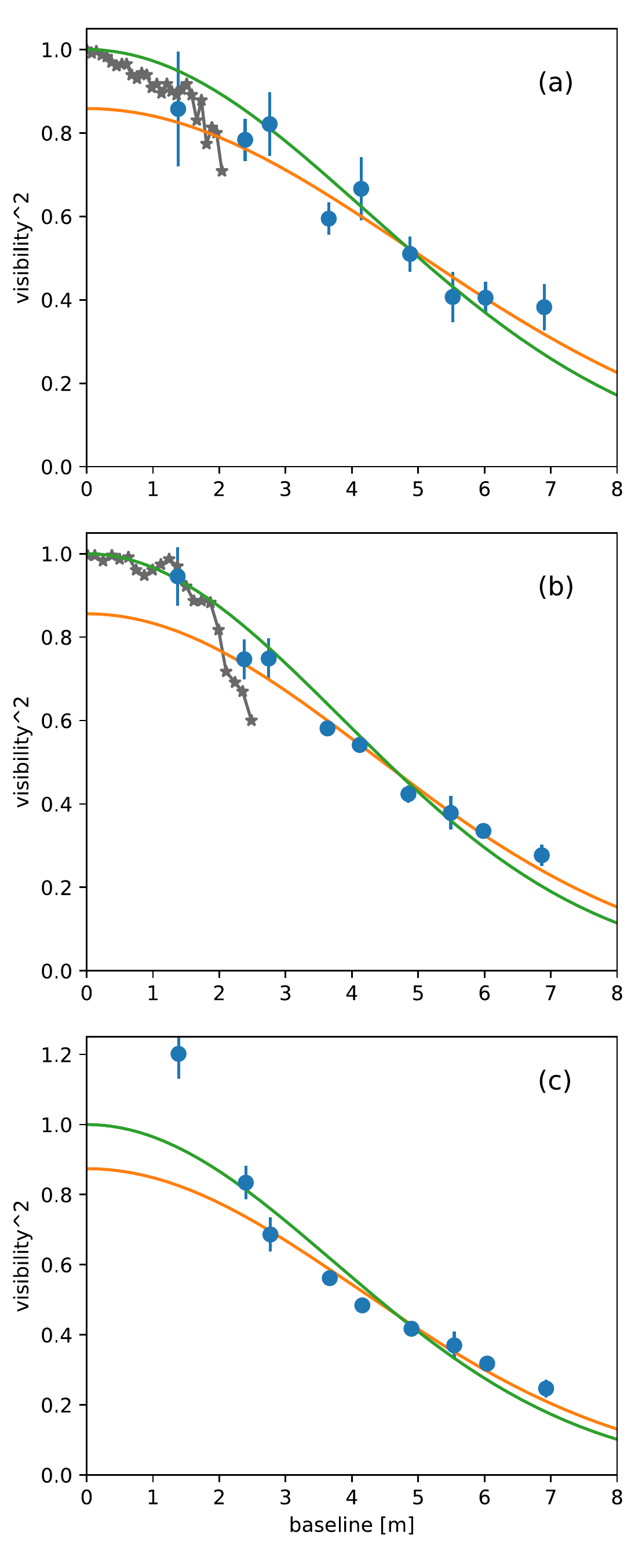}
\caption{The azimuthally averaged SAM visibilities (blue circles) in $Ks$ (a), $L'$ (b) and $M'$ (c). Simple centro-symmetric Gaussial distributions (green and orange lines) have been fitted to the SAM data. The measurements of \citet{lopez1993} are shown for reference (gray stars, cf. Sec.~\ref{comparison}).}\label{comparevislopez}
\end{figure}

\section{Observations and data reduction}\label{obs}

\subsection{NACO}
II~Lup was observed with the NACO instrument on 2010 June 29/30 and July 1 (Prog.ID: 085.D-0356) at the UT4 telescope of the VLT \citep{lenzen2003, rousset2003}. We used the sparse aperture masking mode~\citep[SAM;][]{tuthill2010} where a mask is introduced in the pupil plane and in essence converts the full aperture into a Fizeau interferometer \citep{tuthill2000}. 

One can retrieve information, extending beyond the diffraction-limit of the telescope, from the interference pattern created by the multiple apertures, by measuring the amplitude and phase of the fringes. A large number of apertures allows good Fourier coverage and a high fraction of the phase information recovered by way of the closure phases. An illustration of a typical example of the interference pattern created by SAM is given in Figure~\ref{speckles}.

We have used the 9--hole mask where each hole corresponds to an effective aperture of 0.92 metres. This mask provides 36 baselines with baseline length ranging from 1.3 to 6.9 metres (Fig.~\ref{speckles}). Three broadband filters were used, $Ks$ (2.18 \micron), $L'$ (3.8 \micron) and $M'$ (4.78 \micron), where the angular resolution of the synthesized beam is 32.5, 56.8 and 71.4 mas, respectively. During the first night the seeing fluctuated between 1 and 2.5 arcsec, where the observations in $L'$ and $M'$ suffered from poor seeing. The $Ks$ band observations were carried out during better conditions on the second night. NACO has four different focal cameras and we used the S27 for the $Ks$ band and the L27 camera for the other two broadband filters (for both cases the pixel size is 27.1~mas). We chose a window subframe of $256\times256$ pixels.
 
Two hours were dedicated to the observations of II~Lup and its calibrators. The calibrators measure the telescope transfer function, i.e. the visibility amplitudes of a point source. Thus, the data are calibrated using observations of an unresolved star. The telescope transfer function depends only on the mask and its orientation on the pupil, which remains stable since the pupil-tracking mode was used. In this mode the pupil is kept constant while the sky rotates and the field rotation is later corrected during data reduction. The resolution and the transfer function of the telescope are not dependent on the seeing, but the signal-to-noise is reduced at poorer seeing.

We followed the standard data reduction process for SAM data using a pipeline developed at the University of Sydney \citep{tuthill2000,lacour2011}. For each data cube, the individual speckle frames were flat-fielded, dark, bias, and sky subtracted. In the next step, the frames were Fourier transformed, and the extracted power spectra and bispectra (i.e., the Fourier components) were calibrated. The $L'$ and $M'$ bands data were calibrated using calibrators from both the first and the second night of observations. No significant difference in the telescope transfer function calibration was found overall.

Sparse aperture masking does not provide images, but model images can be derived from the power spectra and the closure phases through image reconstruction methods. We have used two such methods (Sec.~\ref{imaging}), {\sc macim} \citep{ireland2006} and {\sc mira} \citep{thiebaut2008} which differ in the model used for the image reconstruction \citep[see also][for a comparison between the different methods]{lykou2015,beauty2014,beauty2012}.

\subsection{VISIR}\label{visir}
We obtained images of II~Lup in the mid-infrared with the VISIR instrument on the VLT on 2016 March 21\footnote{JD=2457468.366 and $\phi=0.90$, cf. Sec.\ref{comparison}} (Prog. ID: 60.A-9637). The observations were part of the scientific verification program for the instrument's upgrade. We have used the burst mode\footnote{The exposure time was 25ms for PAH\_1 and 27.8ms for PAH\_2.} with a 0.045 arcsec pixel scale, the PAH\_1 ($\rm\lambda_{c}=8.54\,\mu m\,,\Delta\lambda=0.42\,\mu m$) and PAH\_2 ($\rm\lambda_{c}=11.25\,\mu m\,,\Delta\lambda=0.59\,\mu m$) filters, and HD133550 as a point spread function (PSF) calibrator. The chopping and nodding sequence was used to allow for background subtraction and the final reconstruction of images with a shift-and-add technique, i.e. bad pixels were removed from each image and subsequently each image was shifted and added using a maximum of correlation algorithm. The total time on target was 32 minutes. 
A similar process was followed for the PSF calibrator observations. 

The raw images indicated that II~Lup is unresolved in both filters as indicated by the presence of an extended Airy ring pattern, while many images showed instrumental effects due to saturation (`bleeding'). Intensity cuts, bisecting the images north-south and east-west from the central pixel, showed the consecutive Airy rings around the core, while the core's flat plateaus suggest that these images suffer from saturation. We attempted to deconvolve the images using the maximum likelihood method and Lucy-Richardson deconvolution \citep{idlastro}, however this did not improve the images. Therefore, we expect that the circumstellar envelope of II~Lup in the mid-infrared is less than 0.25 arcsec in size. Due to the poor quality of the VISIR images, we opted to not present the raw data, as they are unreliable for any further analysis.

\subsection{ALMA}\label{almareduction}
The ALMA data, from project 2013.1.00070.S, were extracted from the ALMA archive and processed using the reduction scripts from the archive to re-create the fully calibrated measurement set. The ALMA observations were conducted on 2015 August 16, in an array configuration containing 39 of the 12 m antennas with minimum and maximum baselines of 15.06 m and 1574.37 m respectively. This provides a maximum resolution of 0.42 arcsec at frequency $\sim$115.27~GHz (to cover the $^{12}$CO 1--0) transition.

The calibrated data were imaged in {\sc casa} ver.5.1.1-5 \citep{CASA}, with the task {\sc tclean}, using a natural weighting of the visibilities and a $uv$-taper of 0.7 arcsec to taper the signal from the longest baselines. Visibilities from the baselines less than 30 m were also removed during imaging as they contributed a large scale artefact to the final image. The final image created has an RMS noise of 5mJy in a single 2.54\kms\ line-free channel and has a synthesised beam of $1.144\times 0.821$ arcsec$^2$, elongated east-west.
\begin{figure}
\centering
\includegraphics[width=0.45\textwidth, trim=-0.5cm -0.5cm -1cm 0, clip]{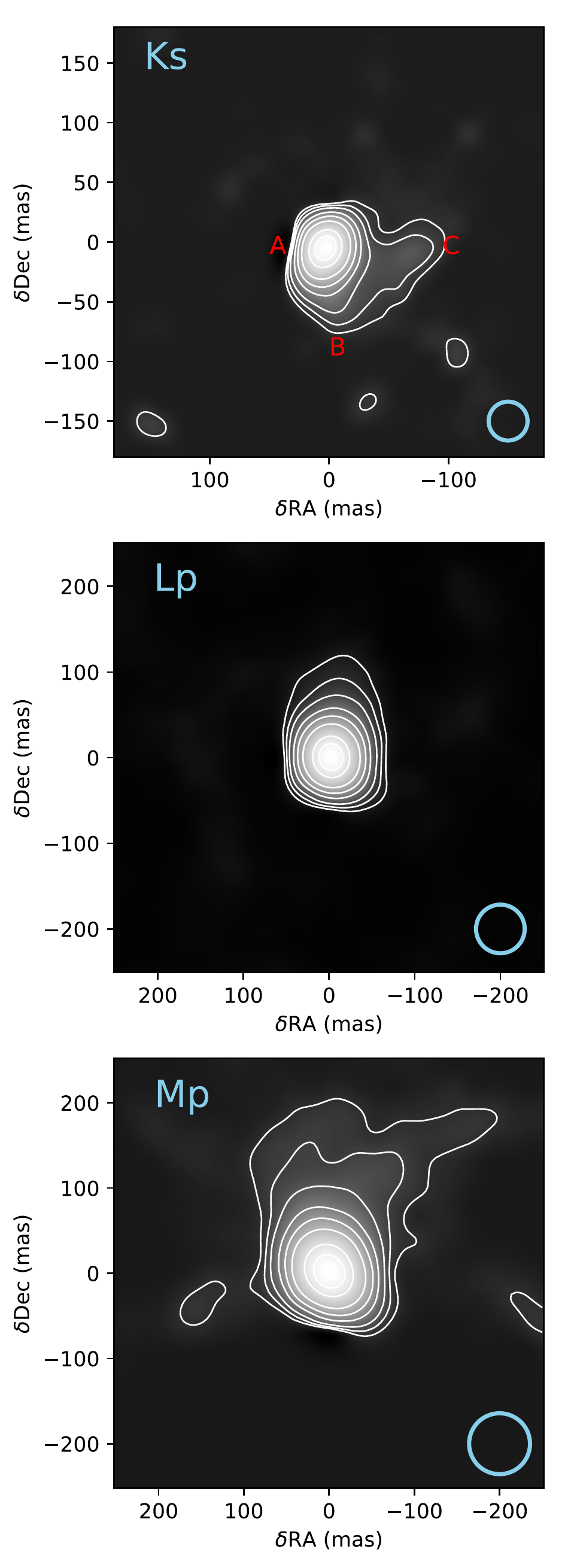}
\caption{{\sc macim} reconstructions for the $Ks$, $L'$ and $M'$ data. Contour levels are
  at 0.5, 1, 2, 5, 10, 20, 50, 75, and 95\% of the peak intensity. The theoretical resolution
  element for each filter is shown in the lower-right corner of each panel. North is up and east
  is on the left. For a discussion on points A to C cf. Sec~\ref{spiralz}.}\label{macimall}
\end{figure}

\begin{table}\label{azimfits}
\centering
\caption{ Equation~\ref{gaussmodel} fitting results.}
\begin{tabular}{ccc}
\hline
filter & $\theta$ (mas) & error (mas) \\
\hline
$Ks$	& 34.4 &2.2	\\
$L'$		& 69.2 &1.5	\\
$M'$	& 90.3 &2.2	\\
\hline
\end{tabular}
\end{table}

\section{Angular sizes}\label{mysizes}

We have azimuthally averaged the SAM data (Fig.~\ref{comparevislopez}) and fitted geometric models to calculate the angular sizes of the dusty envelope at each wavelength \citep[see also][]{lykou2015,sanchez-bermudez2016}. Although this is an oversimplified approach in treating data that sampled different position angles, this method provides initial estimates about the size of the circumstellar envelope that may later be used as {\it a priori} parameters in the image reconstruction process. 


At a first glance (Fig.~\ref{comparevislopez}), the shape of the visibility curve (blue circles) at high spatial frequencies in both $Ks$ and $L'$ suggests that the object may be resolved. We first fit a uniform disc model to the data. The fit was successful for the $Ks$ data, but failed to fit the $L'$ and $M'$ data. As a second step, we fit a Gaussian distribution to the SAM data (orange line, Fig.~\ref{comparevislopez}), whose visibility function , $V(r)$, is expressed as:
\begin{equation}\label{gaussmodel}
V(r)^2 = \left[\dfrac{f_\text{Gaussian}}{f_\text{Gaussian}+ f_\text{background}}\right] \exp\left(-3.57 \theta^2 r^2 \right)
\end{equation}
where $\theta$ is the angular diameter, $r$ is the spatial frequency, coefficient $f_\text{Gaussian}$ is the squared correlated flux at zero baseline, and
$f_\text{Gaussian} + f_\text{background}$ is the total of the correlated and uncorrelated fluxes \citep[the factorial in Eq.\ref{gaussmodel} represents the fractional flux of the envelope; see also similar examples by][]{sanchez-bermudez2016}. The orange curve in Figure~\ref{comparevislopez} represents the fit derived from this scaled model. It is evident that the intercept of the fitted curve at zero baseline is not at $V^2 = 1$, which suggests that the extended flux is over-resolved. For comparison, we show a Gaussian distribution were the coefficient's ratio in Eq.\ref{gaussmodel} was forced to unity and therefore $V^2=1$ at zero baseline (green line, Fig~\ref{comparevislopez}). The angular size of the disc\footnote{The results of the fitted uniform disc and Gaussian distributions were similar for the $Ks$ band.} derived by the model (Eq.~\ref{gaussmodel}) is an upper limit to the real angular size.

The results of the fitting (Eq.\ref{gaussmodel}; Table~\ref{azimfits}) showed that the angular size (Gaussian FWHM) of II~Lup's dusty circumstellar envelope increases at longer wavelengths: $\theta_{Ks}=34.4\pm2.2$ mas, $\theta_{L'}=69.2\pm1.5$ mas, and $\theta_{M'}=90.3\pm2.2$ mas. The observed visibilities suggest a flux contribution by an unresolved point source of approximately 35, 28 and 20\% in $Ks$, $L'$ and $M'$ respectively. However, this fitting procedure is not ideal when the visibilities vary with the azimuth, because essential information is averaged out with this procedure. In that sense, image reconstruction should reveal the true nature of the object.

\subsection{A comparison to Lopez et al.}\label{comparison}

\citet{lopez1993} observed II~Lup in 1991 (JD=2448434, $\phi=0.17$) using one-dimensional ($PA=292.5$\degr) speckle-imaging on the ESO 3.6m telescope in $K$ (2.2\micron) and $L$ (3.6\micron). Their data lacked azimuthal information, but was a first attempt at determining the angular size of II~Lup. The authors modelled the data using a centro-symmetric circumstellar envelope, which provided an adequate fit to the spectral energy distribution of II~Lup, but it could not produce a satisfactory fit for the  visibilities (their fig.~3). The stellar radius and the inner radius of the envelope at 2.2\micron\ were 6~mas and 36~mas, respectively; or 3.5 au and 21 au at our adopted distance of 590 pc.

We have extracted the one-dimensional visibilities of \citet{lopez1993} using the online tool {\sc dexter}\footnote{dc.zah.uni-heidelberg.de/sdexter} and have overplotted them in Fig.~\ref{comparevislopez} for reference (gray stars). The data of Lopez et al. agree in both bands with the SAM data within error bars up to a projected baseline of 2 metre. At 3.6\micron\ the visibilities are lower than those of SAM by a decrement of about 0.15 for projected baselines ranging from 2 to 3 metres. This could be a result of the difference in de-projected baselines, as well as the different broadband filters used.

The angular size we derive for the $Ks$ band is smaller than that of \citet{lopez1993} ($\theta_\text{K, Lopez}=72$~mas), and the estimated unresolved point-source flux contributions differ: 35\% and 28\% (this work), to less than 70\% and 30\% (Lopez et al.) in $K$ and $L$ respectively. This could be explained by (a) the longer baselines ($1.3\leq B_{\rm NACO} \leq 6.9$ to $0\leq B_{\rm Lopez}\leq 3$, in units of metres) and various azimuths offered by non-reduntant aperture masking, (b) the different phase at the time of observation\footnote{NACO observations were obtained at JD=2455377.18. If we adopt $t_0=2447174$ from \citet{lebertre1992} and a period of 575 days, we find that the observations were carried at a phase $\phi=0.266$.}, or (c) a potential change in the morphology of the circumstellar envelope \citep[e.g., IRC+10216,][]{stewart2016A}.

\section{Image reconstruction}\label{imaging}
We present here our image reconstruction results using {\sc macim} and {\sc mira}. We have also tested a third method, {\sc bsmem} \citep{buscher1994}, which gave  results similar to those of {\sc macim}. 

{\sc macim} uses a point source, a uniform disc distribution or a binary as an initial model. We opted for an arbitrary uniform disc model that is unresolved by SAM/NACO with a size of 5~mas, that we assumed is the unresolved stellar diameter. We set the stellar flux contribution at 30--35\% and chose to multiply the visibility errors by a factor of 1.5 and the bispectrum errors by a factor of 2.0 to increase the range and to allow for any miscalibration from the initial data reduction, and taking into account the correlations in the bispectrum data (Ireland, private communication). The $\chi^2$ of the reconstructed maps ranged from 0.83 to 1.0, which suggests that we may have overcompensated for the errors. 

The $Ks$ map of II~Lup is presented in the top panel of Figure~\ref{macimall}. The majority of the emission ($\ge 60$\% of the peak intensity) originates from an area at the centre of the map. At lower emission levels, the shape of the circumstellar envelope is extended within the south-west quadrant and it clearly departs from spherical symmetry. A hook-like extension from the central source is evident. It spreads counter-clockwise from the south and extends West to approximately 100 mas from the origin. The southernmost extension is approximately 80 mas. Although this appears only at low levels of emission (0.5--1\% of the peak), with the exception of three artefacts just south of II~Lup, there is no other point of emission within the map above this level. For the adopted distance of 590 pc, the radius (16 mas) of the unresolved central source translates into 9.44 au and the westernmost extension of the hook translates into 50.15 au.
 
 The morphology of the circumstellar envelope differs at longer wavelengths. The central source is once again unresolved in both maps (middle and bottom panels for the $L'$ and $M'$ maps, respectively, in Fig.~\ref{macimall}). A small protrusion extends at a similar direction ($P.A.\simeq 235$\degr) in both maps. The overall morphology below 50\% of the peak intensity is clearly non-spherically symmetric in both maps, and the shape of the circumstellar envelope is more extended directly north of the central source to approximately 120 and 150 mas at the farthest for the $L'$ and $M'$ maps, respectively. In the $M'$ map, the envelope appears to fan out at low peak-emission levels, however given the poor seeing of the $M'$ observations (cf. Sec.~\ref{obs}), this reconstructed image is of lower fidelity.
 
The hook-like formation seen in the $Ks$ map is absent at longer wavelengths. When the $Ks$ map is convolved with a synthesized beam representative of the $L'$ band observations\footnote{That is, a Gaussian distribution with a FWHM of 60~mas.}, the result indicates that the structure should have been partially resolved in $L'$. Its absence can be explained by the lower resolution and the poorer S/N of that specific night's observations.

\begin{figure}
\centering
\includegraphics[width=\columnwidth]{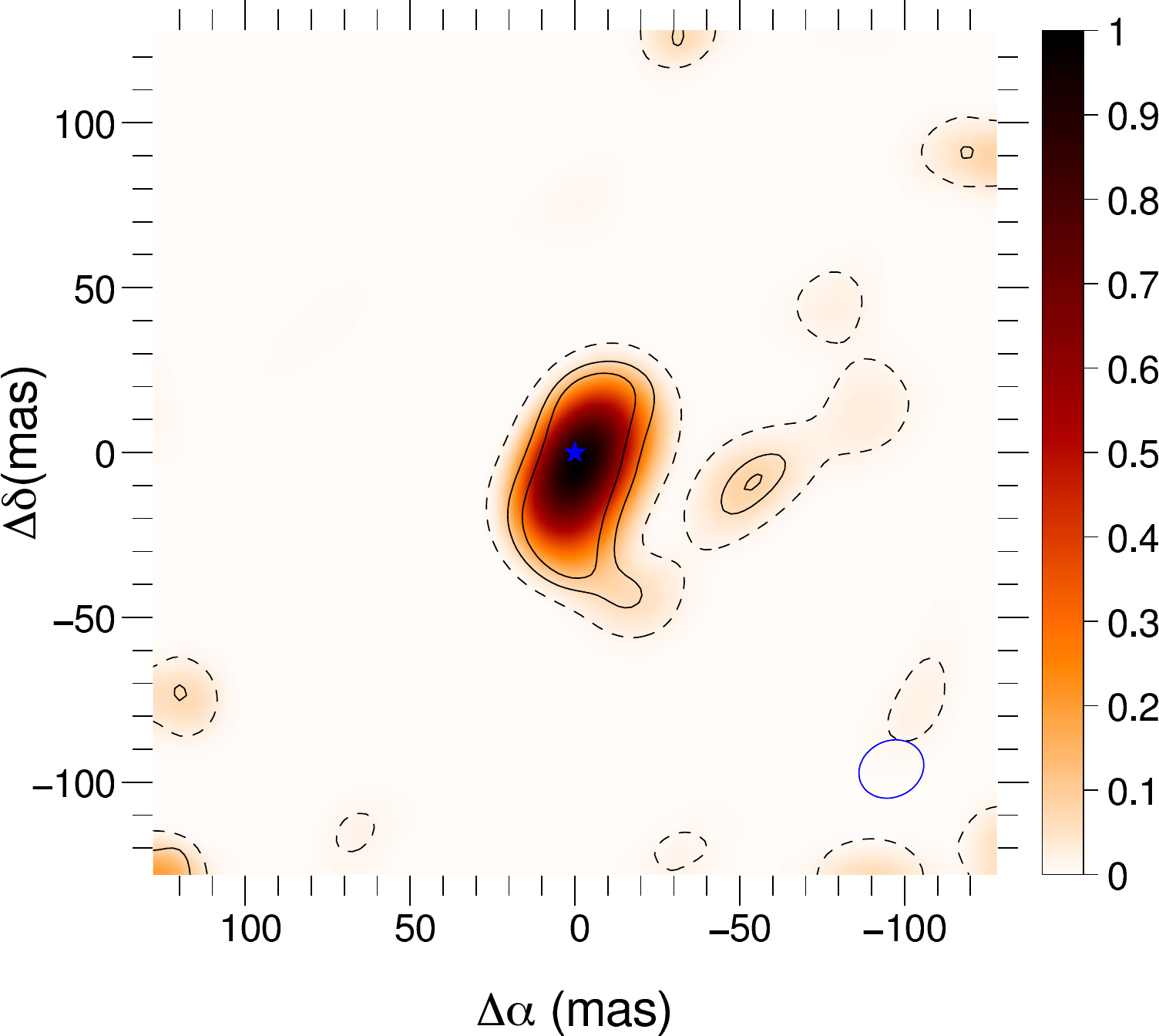}
\caption{{\sc mira} reconstruction for the $Ks$ data. The position of the star is marked by the blue asterisk. The contours are 5, 3 (solid) and 1 sigma (dashed) significance contours. The blue ring indicates the resolution element.}\label{miraKs}
\end{figure}

We used {\sc mira} by incorporating the tool {\sc sparco} \citep{kluska2014} with the same initial parameters as in {\sc macim} for consistency. The quadratic smoothness regularisation of {\sc mira} with a weight $\mu$=1e+9 was used. Figure~\ref{miraKs} shows the resulting image ($\chi^2=0.6$). The overall shape of the structure agrees with the results of {\sc macim}: the elongated envelope and the western clump are reproduced here. At longer wavelengths, the resulting maps were almost identical to those produced by {\sc macim} and therefore are not reproduced here.

\subsection{On the existence of a binary companion}\label{binary}

We now test the hypothesis of \citet{feast2003} that II~Lup may be a member of a binary system. SAM not only allows the recovery of diffraction-limited images, but the Fourier data can also be used to calculate the probability of locating a binary companion. In the past, SAM has enabled many direct detections of planetary companions in the $Ks$ and $L'$ bands \citep[e.g.,][]{lacour2011,gauchet2016}.

We have fitted a binary model to the SAM closure phases (two point sources). The modelled separation ranged between the diffraction limit and the field-of-view. SAM can achieve a contrast ratio up to 250, although the detection limit depends on the brightness of the primary (observed) star\footnote{The use of an aperture mask does decrease the throughput, in this case to $12.1\%$ for the 9-hole mask \citep{tuthill2010}.}. We chose to fit only the $Ks$ dataset because (a) the best angular resolution is offered at that wavelength, and (b) the observations were obtained at better seeing conditions than those at the $L'$ and $M'$ that suffered from poor contrast.

The possible results of the simulations for the $Ks$ dataset indicate the possible presence of a companion within the western tip of the hook-like formation seen in the $Ks$ map (Fig.~\ref{macimall}). The companion is placed at $48.48\,\pm\,5.87$ mas and 262\degr$\pm\,$3.6\degr\ east-of-north from the central source, at a contrast of $29\,\pm\,4$, or a magnitude difference $\Delta Ks=3.64$ ($\chi^2=127.5$). At the adopted distance of 590 pc, the separation would be $28.6\,\pm\,3.46$ au. For this separation and assuming a circular orbit and two solar-mass stars, we derive an orbital period of approximately 108 years, which is significantly larger than the secondary periods derived here and by Feast et al.

This model is not fully convincing, because this companion would require a luminosity one thirtieth of that of the AGB star, or at about 100--200 L$_\text{\sun}$ assuming a similar temperature. A luminous main sequence star can be excluded, as such a star would have evolved faster than II Lup. An evolved red giant is possible but unlikely, because such a companion would have an almost identical initial mass as II Lup. 


If we assume that the 19.7-yr secondary period is the orbital period, then the separation would be 9.19 au, or else 15.6 mas at 590 pc. However, SAM/NACO would not be able to detect such a close companion to II~Lup in $Ks$ band. Such a detection might be possible with the best interferometric instrument, the VLTI, that can theoretically reach an angular resolution of 2 mas at its longest baseline with the auxiliary telescopes (e.g., GRAVITY/VLTI).

\section{Determination of physical properties}\label{physics}

\subsection{Luminosity and effective temperature}
We now attempt to tentatively calculate the physical properties for II~Lup using the primary period, $\log P=2.76$, the bolometric magnitudes found in the literature, and the modelled predictions for the stellar radius given in Table~\ref{envelopes}.

Using the Period-Luminosity relation for carbon-rich Miras of \citet{whitelock2006} ($M_\text{bol}=-2.54\log P+1.87$), we derive an absolute bolometric magnitude, $M_\text{bol}=-5.141$ mag, which consequently provides a luminosity
\begin{equation}
M_\text{bol} - \text{M}_\text{bol, \sun} = -2.5\log(L_*/\text{L}_\text{\sun})
\end{equation}
\noindent of approximately 9800~L$_\text{\sun}$. From this derivation and the theoretical stellar radii, we calculate the effective temperatures, $T_{\rm eff}$, given in Table~\ref{physical}. Although our derived luminosity differs from the models found in the literature, the range of the derived effective temperatures is congruent with the range of the modelled effective temperatures \citep[2200--2800 K;][]{danilovich2015,ramstedt2014,debeck2010,dehaes2007,schoeier2007,ryde1999,groenewegen1998}. The only exception is the effective temperature (2082~K) derived for the stellar radius\footnote{The angular radius of 6~mas projected at a distance of 590 pc.} of \citet{lopez1993}; the temperature adopted by Lopez et al. was 2200 K.

\subsection{Dust temperature}
As a comparison, we estimate the temperature of the circumstellar envelope in the $Ks$ band using the angular size derived from our observations, $\theta_{Ks}=34.4\,\pm\,2.2$~mas, the apparent bolometric magnitude $m_\text{bol}=3.71$, and the relation of \citet{bedding1997},
\begin{equation}
\log T = 4.22 -0.1m_\text{bol} -0.5\log \theta_{Ks}
\end{equation}
\noindent Although the relation above should provide the effective temperature of a star, the angular size derived from the SAM/NACO observations does not correspond to the stellar diameter. The derived temperature, $T\sim1200$~K, is that of the dust enclosed within said angular size. 

\begin{figure}
\centering
\includegraphics[width=0.48\textwidth]{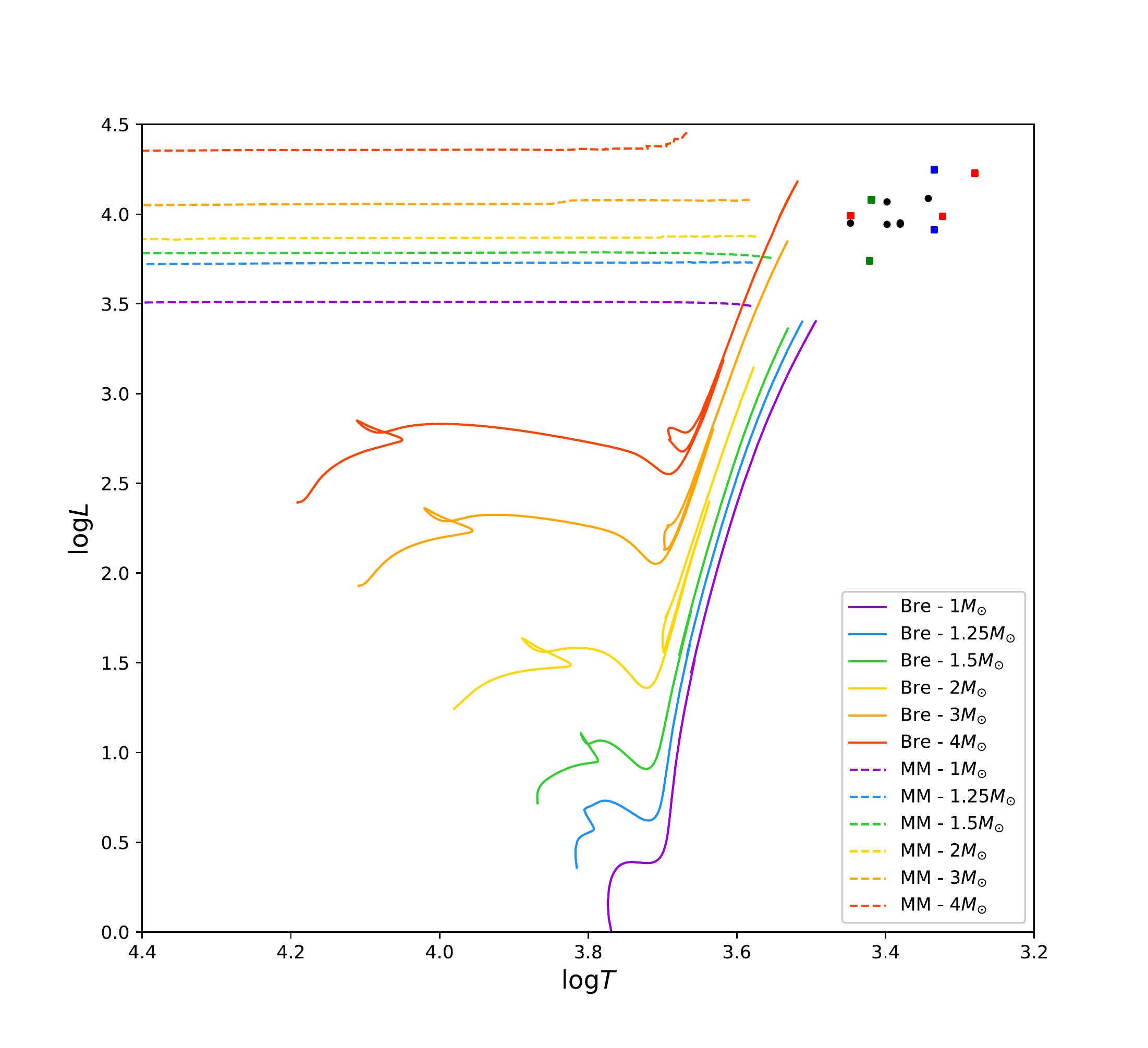}
\caption{Evolutionary tracks of intermediate-mass stars ($1-4$~M$_\text{\sun}$) from \citet{bressan2012} (bold lines) and from \citet{millerbertolami2016} (dashed lines). The physical parameters of II~Lup ($\log T$, $\log L$; black circles) were extracted from the theoretical models found in the literature \citep{danilovich2015,ramstedt2014,schoeier2013,debeck2010,dehaes2007,groenewegen1998,lopez1993}. Three additional carbon-rich AGB stars are overplotted on the H--R diagrams for comparison: R~Scl \citep[green squares,][]{wittkowski2017,bergeat2002}, V~Hya \citep[blue squares,][]{groenewegen2002b,bergeat2002} and CW~Leo \citep[red squares,][]{ramstedt2014,groenewegen2002b,bergeat2002}. For a further description, see main text (cf. Sec.~\ref{calcmass}).}
\label{hrdiag}
\end{figure}

\subsection{Mass}\label{calcmass}
\citet{ryde1999} have characterized II~Lup as a `highly evolved carbon star' with a significantly lower carbon isotopic ratio \citep[$\rm 4\leq{}^{12}C/^{13}C\leq 10$ , see also][]{woods2003,debeck2010,ramstedt2014} as opposed to typical carbon stars  \citep[$\sim 30$;][]{nyman1993,greaves1997}. According to the authors, the lower isotopic ratio could suggest that II~Lup may have been a $5-8$~M$_\text{\sun}$ star, where hot-bottom burning has been suppressed therefore preventing the star from becoming oxygen-rich. Here, we use three different methods to estimate the mass of II~Lup based on the theoretical stellar radii in Table~\ref{envelopes}. These methods are described below.

\subsubsection{Evolutionary tracks}
We attempt to place II~Lup on a Hertzsprung-Russell diagram. We selected evolutionary tracks for stars of solar metallicity and with initial masses between $1-4$~M$_\text{\sun}$ from \citet{bressan2012} and from \citet{millerbertolami2016} to cover the evolutionary sequences until and beyond the AGB phase, respectively. These evolutionary tracks are shown in Figure~\ref{hrdiag}. The physical parameters of II~Lup ($\log T$, $\log L$; black circles) were extracted from the models found in the literature \citep{danilovich2015,ramstedt2014,schoeier2013,debeck2010,dehaes2007,groenewegen1998,lopez1993}. As a comparison, we show the locations of three other carbon-rich AGB stars on these H--R diagram: R~Scl \citep[green squares,][]{wittkowski2017,bergeat2002}, V~Hya \citep[blue squares,][]{groenewegen2002b,bergeat2002} and CW~Leo \citep[red squares,][]{ramstedt2014,groenewegen2002b,bergeat2002}. Despite the uncertainties in the physical properties of the theoretical models, the H--R diagram indicates that II~Lup is placed near the tip of the AGB. We could therefore deduce that the initial stellar mass is $1.5\pm0.5$~M$_\text{\sun}$, however we must point out that the \citet{bressan2012} evolutionary tracks terminate at the tip of the RGB for stars below 2~M$_\text{\sun}$ and at the beginning of the TP-AGB phase at higher masses.

\subsubsection{The method of \citet{fox1982}}
Following the method described in \citet{soszynski2013b}, we calculated the extinction-corrected Wesenheit index in the near-infrared, $W_\text{JK}$, using the the 2MASS photometry and the distance of \citet{pietrzynski2013} for the Large Magellanic Cloud. The resulting value ($W_\text{JK}=8.53\pm0.05$) places II~Lup near the tip of the fundamental mode of carbon-rich stars in Figure 1 of \citet{soszynski2013b}.

Therefore, if we assume that the primary period of II~Lup is on the fundamental mode, and following the pulsation equation of \citet{fox1982}, the stellar mass is given by
\begin{equation}
M = \left(\frac{Q}{P} \right)^2 R^3
\end{equation} 
\noindent where $M$ and $R$ are given in solar units, and $Q\sim0.1$ for the fundamental mode \citep[based on the predictions for $0.8\lesssim M\lesssim3.0$~M$_\text{\sun}$ of][]{fox1982}. Using this formula we can tentatively estimate the mass of II~Lup for the given stellar radii (Table~\ref{physical}). The range of the derived values is quite large, however the calculations were based on parameters with large uncertainties, such as the theoretical stellar radii and the distance.

\begin{table}
\centering
\caption{Physical properties of II~Lup}\label{physical}
\begin{tabular}{ccrl}\hline
$R_*$~(au) & $T_\text{eff}$~(K) & \multicolumn{2}{c}{$M$~(M$_\text{\sun}$)} \\
 & & (1) & (2)\\ \hline
1.86 & 2869& 1.95 & 1.87\\
2.32 & 2571& 3.76 & 2.99\\
2.52 & 2466& 4.83 & 3.59\\
2.54 & 2456& 4.95 & 3.66\\
2.66 & 2399& 5.71 & 4.05\\
3.00 & 2246& 8.46 & 5.38\\
3.54 & 2082& 13.34 & 7.48\\
\hline
\multicolumn{4}{l}{\textbf{Methods:} (1) \citet{fox1982}; }\\
\multicolumn{4}{l}{(2) \citet{vanleeuwen1997}}\\
\end{tabular}
\end{table}

\subsubsection{The method of \citet{vanleeuwen1997}}
We now attempt to estimate the stellar mass using the relation for the fundamental mode of \citet{vanleeuwen1997}:
\begin{equation}
\log M = (1.949\log R -2.07 -\log P)/0.9
\end{equation}
\noindent and the resulting values are tabulated in Table~\ref{physical}.

The initial mass of a carbon-rich AGB star can be within the range of $1-4$~M$_\text{\sun}$, but this is highly dependent on the metallicity \citep{karakas2016,millerbertolami2016}. However, the stellar masses derived by the method of \citet{fox1982} for the given stellar radii vary by a large factor. The last two values are much higher than the mass limit for intermediate-mass stars. Therefore, only theoretical radii below 2.5 au would provide realistic stellar masses for II~Lup based on the methods of \citet{fox1982} and \citet{vanleeuwen1997}, i.e. $\la3.6$~M$_\text{\sun}$.

\begin{figure*}
\centering
\includegraphics[width=0.9\textwidth, trim= 0cm 0cm 0cm -0.3cm,clip=True]{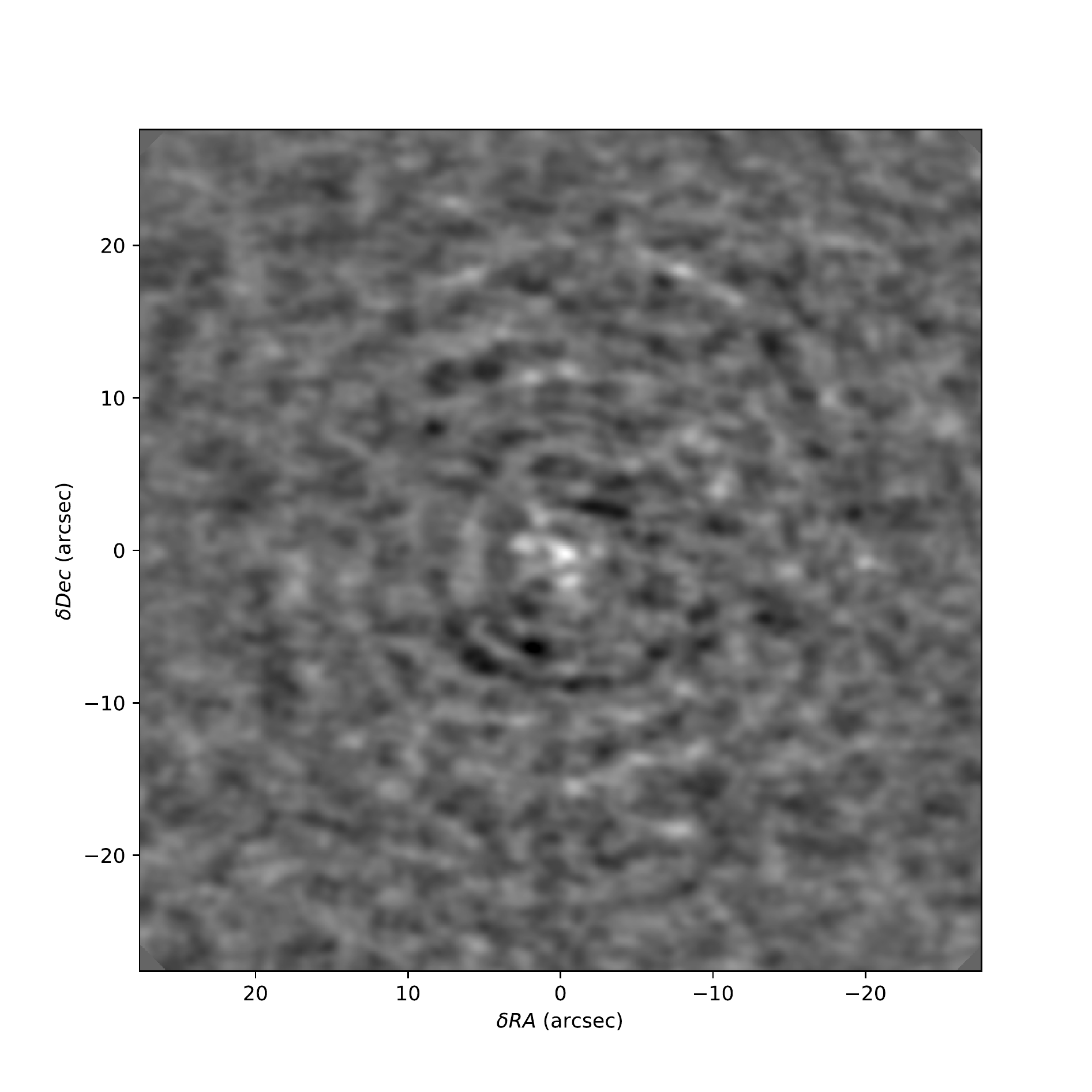}
\caption{ALMA CO ($J=1-0$) channel map. The large scale structure extends approximately 23 arcsec from the central star.}
\label{COspirals}
\end{figure*}

\begin{figure*}
\centering
\includegraphics[width=\textwidth]{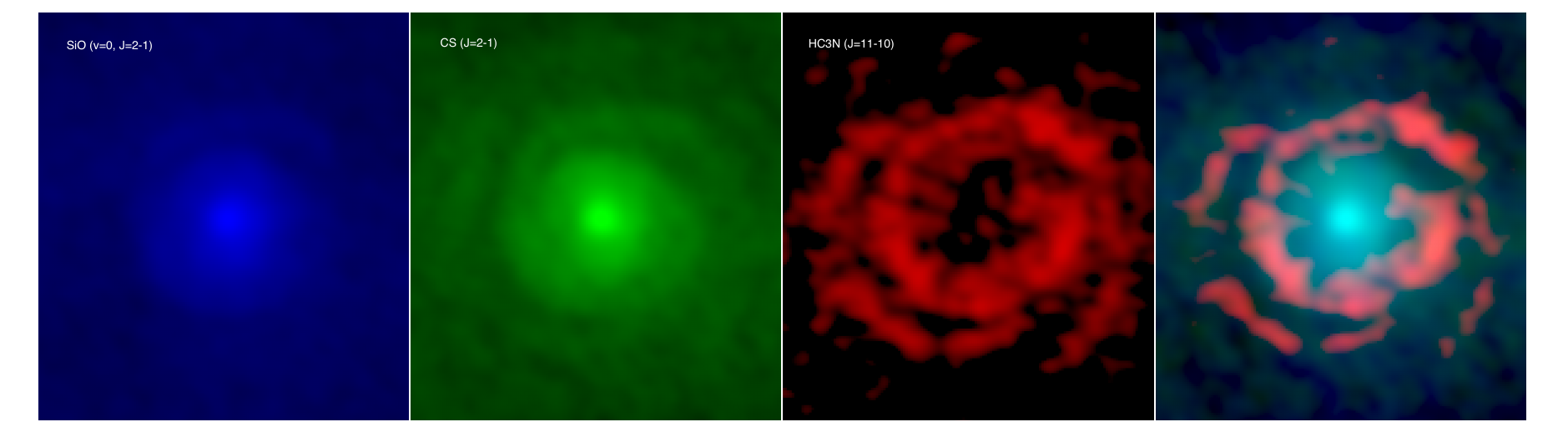}
\caption{The inner coils of II Lup's spiral shown in three different molecules, namely SiO, CS and HC$_3$N from left to right. The last panel shows a false-colour image of the spiral based on these three molecules. North is up and east is on the left.}\label{colourspiral}
\end{figure*}
\section{The circumstellar environment of II Lup}\label{discussion}

\subsection{On a putative spiral formation detected by NACO}\label{spiralz}

Let us now consider the timescale of the expansion of the structure in the $Ks$ map (Fig.~\ref{macimall}). We adopt the gas expansion velocity $v_\text{exp}=22.2$\kms \citep{nyman1993}, and assume that the dust velocity is similar and that the expansion is linear. We select three locations at the 0.5\% peak-emission level on the $Ks$ map (Fig.~\ref{macimall}): one directly east of the central source (hereby point A), one directly south (point B), and one directly west (point C). Following our previous assumptions and again adopting a distance of 590 pc, the timescales for the expansion of the dust up to points A, B and C are 4, 8.9 and 11.7 years respectively. However, if the ejecta were expanding radially and the expansion was spherically symmetric, the ejecta would not have been located only in the lower-right quadrant of the map. 

We now assume that the counter-clockwise hook-like extension in the $Ks$ map of II~Lup (Fig.~\ref{macimall}) is the inner arm of a spiral. Spiral formations are not uncommon, however only a handful of spirals have been detected around AGB stars in the last decade. These characteristic examples are: AFGL~3068 \citep{kim2017,mauron2006}, R~Scl \citep{maercker2012}, RW~LMi \citep{monnier2000,kim2013}, and CW~Leo \citep{decin2015}. All four stars are carbon-rich like II~Lup. Observations suggest that two oxygen-rich stars, Mira \citep{ramstedt2014a} and W Aql \citep{mayer2013}, and one S-type star \citep[the possibly triple system of $\pi^1$~Gru,][]{mayer2014}, also indicate the presence of a spiral. All of the examples mentioned above are known binaries or strongly expected to belong in a binary system.


Following these assumptions, if we consider that the coil of the putative spiral reached the western tip of the hook (point C) within 12 years, it could complete a counter-clockwise rotation over four quadrants within 40 years. However, the full extent of the influence of binarity in such spirals is seen at larger spatial scales ($\ga10$ arcsec) compared to the field-of-view of SAM/NACO. We cannot exclude a longer period considering many AGB stars with spirals have longer orbital periods (e.g., R Scl: 350 years \citep{maercker2012}, AFGL 3068: 830 years \citep{mauron2006}), and we demonstrate this argument in the following sections.


\begin{table}
\centering
\caption{ALMA results}\label{almainfo}
\begin{tabular}{llcc}\hline
Molecule & Transition & Frequency (GHz) & $D^\ast$ (arcsec)\\
\hline
$^{12}$CO & $J=1-0$ & 115.276 & $\la56$ \\
HC$_3$N & $J=11-10$ & 100.083 & $\sim14.2$ \\
CS & $J=2-1$ & 97.9873 & $\sim10.4$ \\
SiO & $\nu=0, J=2-1$ & 86.8526 & $\sim10$ \\
\hline
\multicolumn{4}{l}{($\ast$): diameter of the circumstellar envelope.}\\
\end{tabular}
\end{table}

\begin{figure}
\centering
\includegraphics[width=\columnwidth, trim=0cm 0cm 0cm 0.05cm,clip]{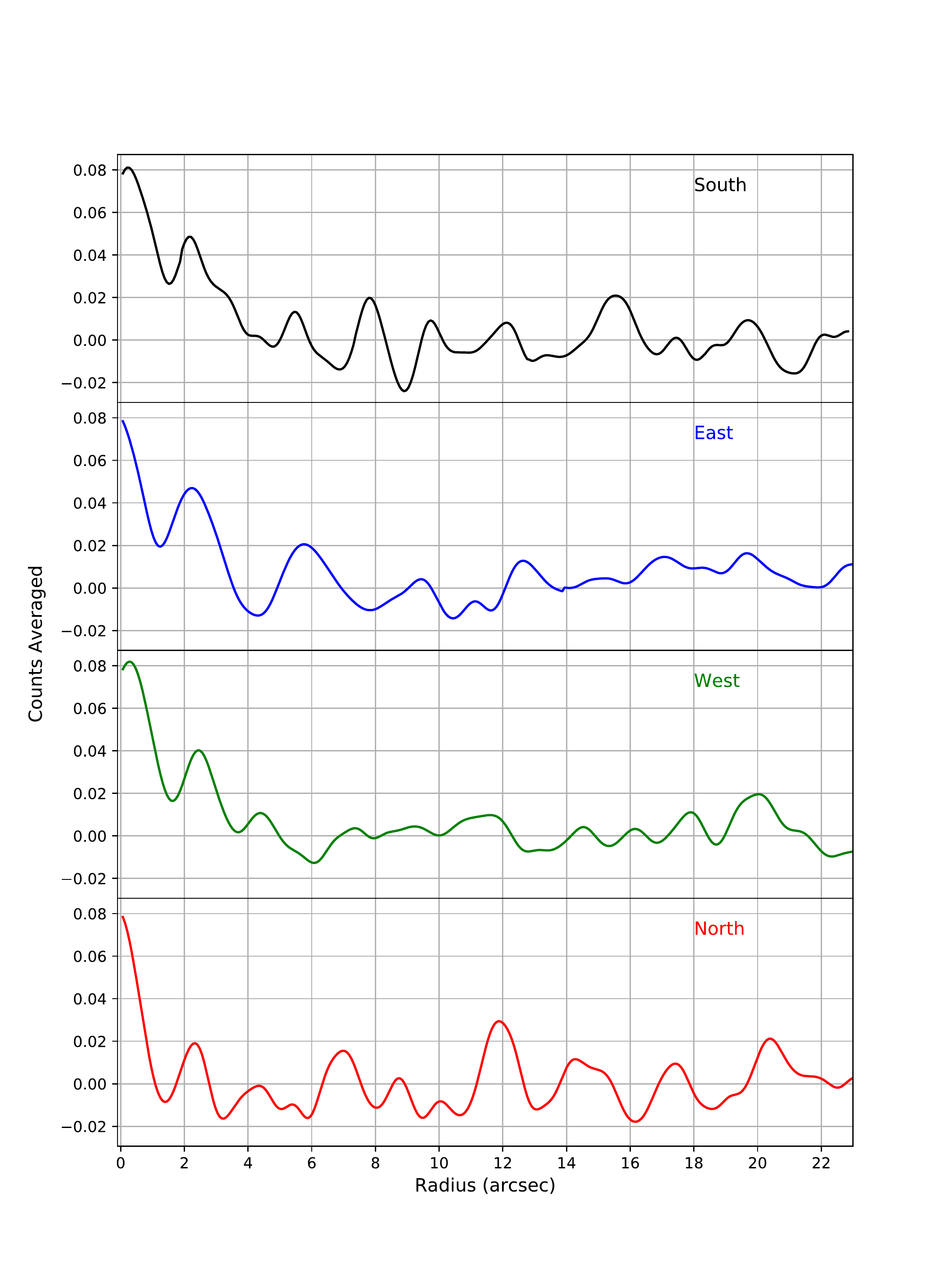}
\caption{Radial profile projection from the position of II~Lup (2MASS coordinates) and extending up to 23 arcsec in all cardinal points to illustrate the location of the spiral arms in each direction.}\label{cocuts}
\end{figure}

\begin{figure}
\centering
\includegraphics[width=\columnwidth]{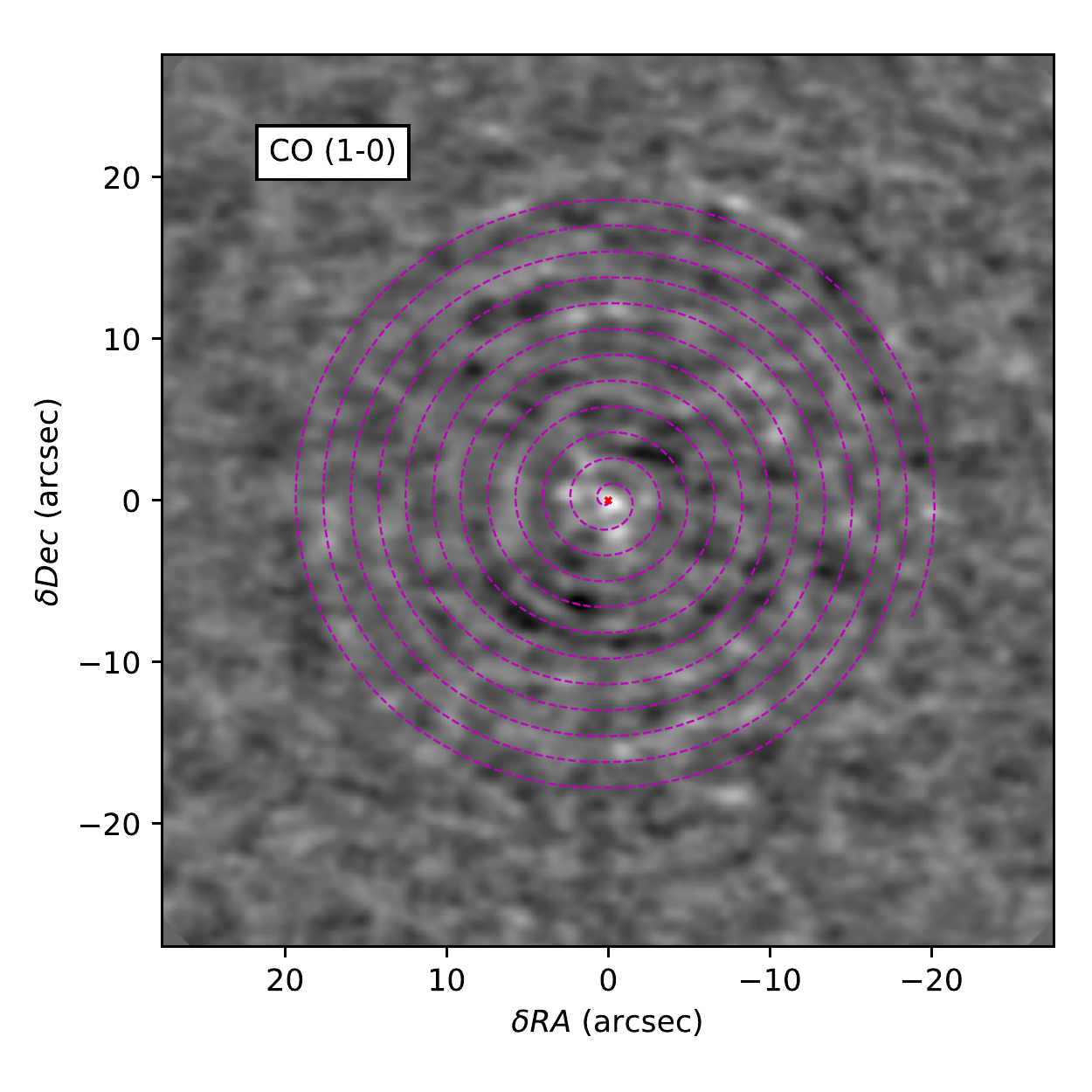}
\caption{A counter-clockwise spiral with a 1.7 arcsec arm separation overplotted on the CO channel map of Fig.\ref{COspirals}. The red star marks the position of II Lup in the {\it Gaia} system.}
\label{traces}
\end{figure}

\subsection{Tracing the circumstellar envelope with ALMA}\label{alma}

To test the validity of the asymmetries detected with aperture masking, we retrieved an ALMA data set from the archive (cf. Sec.~\ref{almareduction}). The observations covered a number of spectral lines, at sub-arcsecond resolution.
Images of four detected emission lines are shown in Figures \ref{COspirals} and \ref{colourspiral}. The lines are from the molecules CO, SiO, CS, and HC$_3$N (Table~\ref{almainfo}). SiO and CS are detected from the innermost region: these are molecules that exist in the inner wind. HC$_3$N forms a little further out and traces the inner envelope but not the near-stellar region. CO is detected throughout the envelope. The images show the spectral channel corresponding to the central velocity.

We find that the circumstellar, molecular envelope of II Lup extends as far out as 23 arcsec from the central source in the CO (1--0) transition. The most striking discovery from this dataset is a spiral traced by the carbon monoxide gas (Fig. \ref{COspirals}). The spiral arms appear patchy and faint at places, but the overall structure is evident in the CO map. The ALMA CO  image is significantly affected by missing short spacings. It is therefore sensitive to small-scale structure which is dominated by the spiral pattern, but the large structure is resolved out.



The last panel of Fig.\ref{colourspiral} shows a false-colour image composed of three molecules (SiO, CS, and HC$_3$N). Only one northern coil extending clock-wise can be seen in the SiO and CS channel maps, while the overall emission is dominated by the central core. The coil is an exact match in both molecules. On the other hand, the HC$_3$N channel map shows the absence of a central source, thus indicating that the molecular emission arises only from the circumstellar envelope. The most striking feature in the HC$_3$N map is the distinct separation of the structure in two different arms.

All four images show a clear spiral signature. The location of the windings depends on the velocity: the selected channels are those where the windings are located furthest out, as expected at the systemic velocity for an expanding envelope. Because the observed location changes with the selected velocity, averaging of spectral channel reduces the contrast of the spiral.

From the position of II Lup (based on the 2MASS coordinates), we obtain radial cuts in the CO image in four directions: south, north, east and west. The profiles are shown in Fig.~\ref{cocuts}. The profiles show clear peaks where the cuts crosses a spiral winding: a total of 12 windings were detected, out to the largest distance of 23 arcsec. The peaks were measured to obtain their distance from the star. We measured the separations for each of these peaks. For each cut, we calculated the median separation: 1.565, 1.595, 1.69 and 1.75 arcsec, for south, north, east and west, respectively. The separation is slightly larger in the east-west direction than in the north-south, suggesting there may be some flattening of the spiral. However this difference is marginal. We find a median separation of 1.7 arcsec. A toy-model spiral with this separation is overplotted on the CO map in Fig.\ref{traces}.


\begin{figure}
	\centering
		\includegraphics[width=\columnwidth]{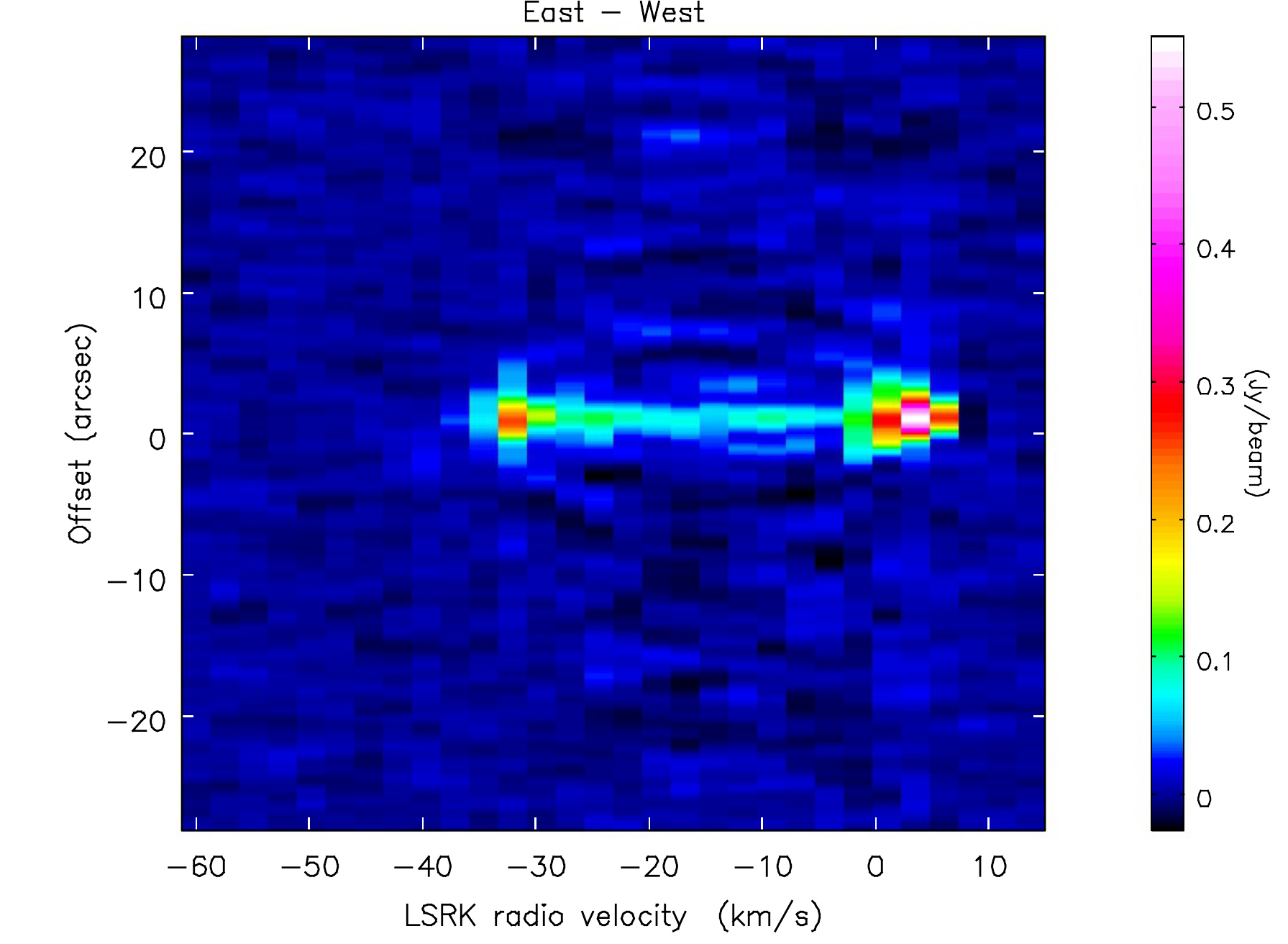}\\
		\includegraphics[width=\columnwidth]{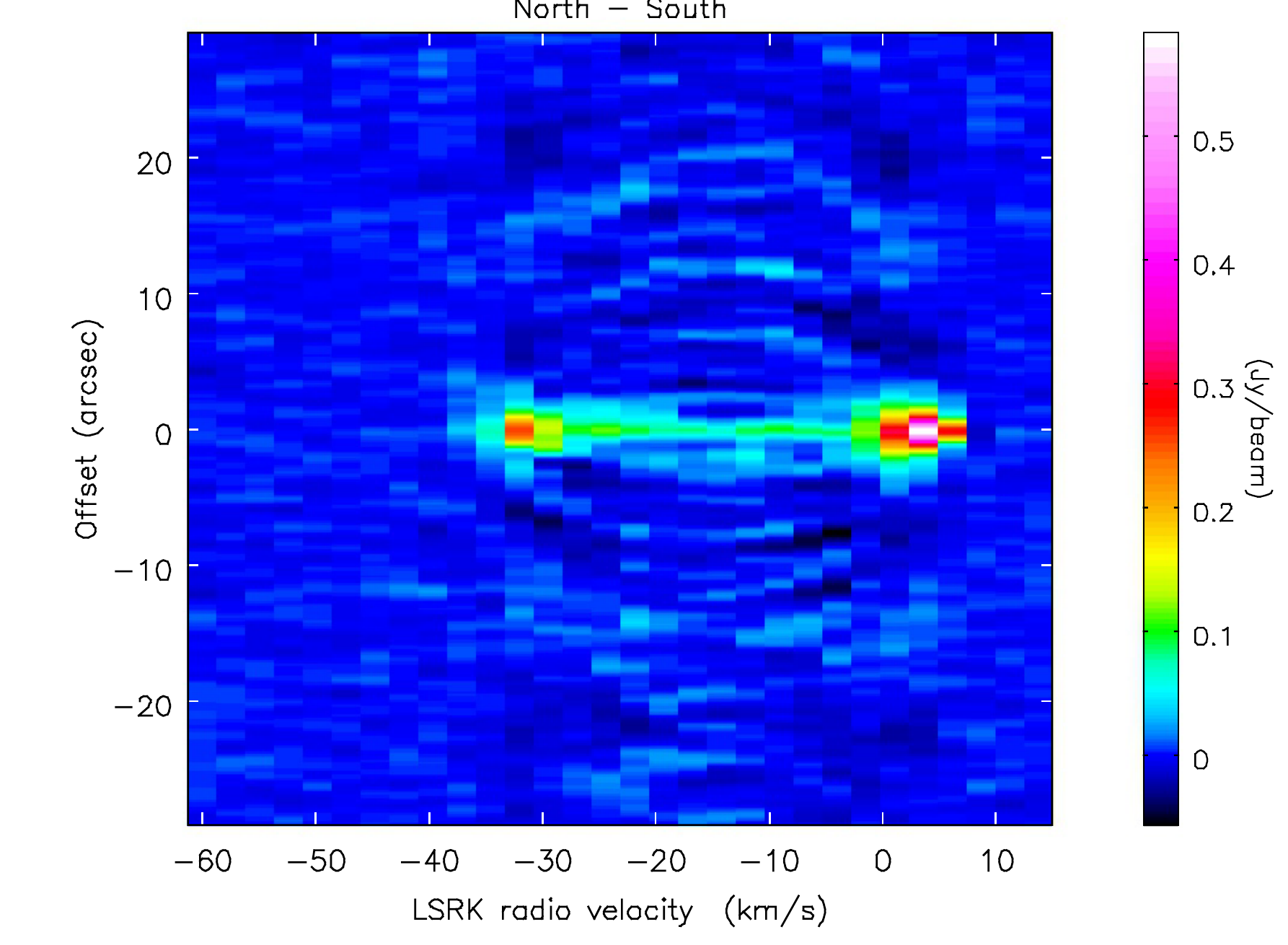}
	\caption{Position-velocity diagrams along the east-west (top) and north-south axis (bottom) through the centre of the CO map.}
	\label{pvdiagrams}
\end{figure}

Figure~\ref{pvdiagrams} presents the position-velocity (P-V) diagrams along the north-south and east-west axis of the CO (1--0) map (Fig.~\ref{COspirals}). At a first inspection, the P-V diagrams show a narrow bar of emission at the centre with enhanced features at $+5$\kms\ and $-35$\kms\ corresponding to the core of Fig.~\ref{COspirals}. At lower intensity levels (below 0.1 Jy/beam) arc features are visible\footnote{For a better view we refer to the electronic version of this figure.} extending up to about $\pm20$ arcsec along the north-south axis (bottom panel, Fig.~\ref{pvdiagrams}). These features are less prominent in the east-west P-V diagram (top panel, Fig.~\ref{pvdiagrams}), although two arcs are visible within $\pm10$ arcsec from the central bar. 

We compare these diagrams to the work of \citet{homan2015}, who simulated CO emission from the circumstellar envelopes of AGB stars, and in particular to their modelled P-V diagrams. We find that the P-V diagrams of Figure~\ref{pvdiagrams} (this work) are somewhat similar to both those in Fig. 9 of \citet{homan2015} for their model of an edge-on, narrow spiral with a low mass-loss wind and a more massive homogeneous-to-spiral wind outflow, as well as those in their fig. 11 for a `shell spiral' with a low mass-loss wind and an equal homogeneous-to-spiral wind outflow\footnote{\citet{homan2015} distinguish a `narrow' from a `shell spiral' depending on their definition of the spiral's opening angle, that is how their model spirals open with the respect to the orbital plane. A small opening angle results in a `narrow' spiral, while an opening angle of 180\degr\ gives a `shell spiral'. For a more in-depth description, we refer to \citet{homan2015} and their sect.2.1.2.  }. As shown here in Fig.~\ref{COspirals},  the structure does not appear to be edge-on and the east-west P-V diagram in Fig.\ref{pvdiagrams} is not `S-shaped', therefore the first case of \citet{homan2015} for the narrow spiral might not apply here. The models of \citet{homan2015} for their `shell spiral' indicate slightly similar P-V diagrams with a bright, narrow strip at the centre of the P-V diagram (their fig.11) that is not seen here (this work, Fig.~\ref{pvdiagrams}). 

This is a tentative comparison since in the work of \citet{homan2015} the P-V diagrams were modelled for a star similar to CW~Leo and for the CO $J=3-2$ transition which traces the inner wind unlike the CO $J=1-0$ transition. Moreover, the mass loss of II~Lup is significantly higher than the one in the low mass-loss wind model of \citet{homan2015}. As a last point, the simulated observations of \citet{homan2015} for the ALMA C34-7 configuration (their fig.19, panel D) does not match the observations presented here that were obtained with a similar configuration in 2015. Further observations will be required to re-visit the complexity of the circumstellar envelope of II Lup as seen by ALMA.

As a last step, we follow the analysis of \citet{mayer2013} to determine initial parameters for the putative spiral around II~Lup, and use their equations for the arm separation, 
\begin{equation}
\rho=v_\text{exp} \cdot P \cdot \upi
\end{equation}
\noindent where here $\rho$ is the arm separation given above, $P$ is the orbital period and $\upi$ is the parallax (for the adopted distance of 590 pc). We derive an orbital period of about 128 years.

\subsection{Comparing the results of SAM/NACO and ALMA.}

Figure~\ref{traces} shows the spiral with the 1.7 arcsec arm separation superposed on the CO image. The starting angle of the spiral was used as a free parameter. The spiral was given the slight flattening indicated by the CO data. It provides an acceptable fit. There is also some complexity which goes beyond a simple spiral structure. This is expected from models \citep{mohamed2012} which show that in binary motion, there is a main spiral caused by the movement of the centre of mass loss, but there is a second, counter spiral caused by interaction between the wind and the companion.

It is possible that the secondary arm, or a portion of it near the mass-losing star, was detected by SAM/NACO observations. In Figures~\ref{macimall} and~\ref{miraKs}, the envelope is shown to extend south-west within $190^o \la$P.A.$\la 235^o$, which is not clearly aligned with the spiral traced by the molecular gas. The models of \citet{mohamed2012} suggest no structure inside the stellar orbit, which would imply that the binary companion should not be further out than component `C' ($Ks$ map of Fig.~\ref{macimall}). It is of course still possible that component `C' is the binary, or at least emission associated with it. In that case, the luminosity of the companion ought to be less than 200~L$_\text{\sun}$ or else it would have been clearly detected by SAM/NACO. If we assume that the initial mass of II~Lup was within $2.5\pm1$~M$_\text{\sun}$, as calculated in Sec.\ref{calcmass}, then a 128-year period would imply $m_2\la1$~M$_\text{\sun}$. Since the detected structure appears to be seen nearly pole-on, this hypothesis can be tested by further observations in the near infrared.

The indication of a larger separation would be consistent with a reduction in the mass of the primary star or an increase in the wind speed. However, the indication is only tentative. We find that the ALMA data provides strong support for the existence of a spiral structure, which implies that binarity is the main shaping mechanism of the mass loss, therefore the effects of a companion's orbital motion is the most likely explanation of the asymmetries seen in the aperture masking data.

\section{Conclusions}\label{conclusions}

The carbon star II~Lup is known to exhibit an obscuration event in addition to a primary period of 575 days. According to \citet{feast2003} this can be explained either by the ejection of dust clumps or by the presence of a companion. After re-analysing the infrared lightcurve of II~Lup, we confirm a secondary period of approximately 19 years. However, due to the incompleteness of the sample, the true secondary period may be significantly longer. We attempted to estimate the physical parameters of II~Lup (luminosity, effective temperature and mass), however we find that the results are inconclusive due to the large uncertainties of the measurements used. Nevertheless, we obtain a dust temperature of $\sim$1200 K from the $Ks$ angular size of II~Lup. 

This work presents the first-ever images of the dusty envelope of II~Lup in near-infrared and sub-mm wavelengths. We have used an aperture masking technique to obtain diffraction limited images from a single-dish telescope in $Ks$, $L'$ and $M'$. The interferometric data revealed that: 
\begin{itemize}
\item[(i)] the angular size of the source, $\theta_{Ks}$, is less than 35 mas, and therefore the stellar radius is much smaller than 20 au (assuming a distance of 590pc), 
\item[(ii)] the morphology of the circumstellar envelope is non-spherical and that its shape is hook-like at low emission levels ($Ks$ map, Fig.~\ref{macimall}), possibly resembling the inner coil of a spiral, or associated with a counter-spiral shaped by orbital motion \citep[e.g., ][]{mohamed2012},
\item[(iii)] the morphology of the envelope is different at longer wavelengths ($L'$ and $M'$), where the envelope expands directly north from the central source. However, due to a lower resolution and signal-to-noise at those wavelengths, the fidelity of those images is lower and we can conclude that the envelope is smaller than 200 mas in these bands.
\end{itemize}

Our analysis of the archived ALMA channel maps from $^{12}$CO ($J=1-0$), SiO ($\nu=0, J=1-0$), CS ($J=2-1$) and HC$_3$N ($J=11-10$), reveals the true extent of the non-spherical circumstellar envelope and most importantly hints at a large-scale spiral structure in all maps. From the spacing of the spiral arms in the CO map ($\sim1.7$ arcsec), we derive an approximate orbital period of 128 years. A study of the higher CO transitions tracing the inner wind of II Lup would be beneficial in determining the history of the mass loss process.

Given the complexity of this object we suggest new observations of the circumstellar environment of II~Lup in order to compare the morphology at different epochs. We strongly recommend (a) further monitoring of its variability in the visual and near-infrared, and (b) a polarimetric and spatio-kinematic study of its non-spherical dusty environment with large-aperture telescopes (e.g., GRAVITY/VLTI and SPHERE/VLT).




\section*{Acknowledgements}
This research is based on observations collected at the European Southern Observatory under ESO programmes 085.D-0356(A) and 60.A-9637(A). We thank the European Southern Observatory for allocating time on NACO and VISIR for this project. We acknowledge with thanks the assistance of the NACO team during the challenging observations. 

We thank Patrick Lenz for his assistance with {\sc period04}. FL acknowledges support from the Hong Kong University Postdoctoral Fellowships scheme and the FWF AP23006 project (PI: Josef Hron). JK acknowledges support from the Philip Leverhulme Prize (PLP-2013-110, PI: Stefan Kraus), and from the research council of the KU Leuven under grant number C14/17/082.  This paper makes use of the following ALMA data: ADS/JAO.ALMA\#2013.1.00070.S. ALMA is a partnership of ESO (representing its member states), NSF (USA) and NINS (Japan), together with NRC (Canada), MOST and ASIAA (Taiwan), and KASI (Republic of Korea), in cooperation with the Republic of Chile. The Joint ALMA Observatory is operated by ESO, AUI/NRAO and NAOJ. Based on observations obtained with {\em Planck} (http://www.esa.int/Planck), an ESA science mission with instruments and contributions directly funded by ESA Member States, NASA, and Canada. {\em Herschel} is an ESA space observatory with science instruments provided by European-led Principal Investigator consortia and with important participation from NASA.

This research has made use of {\sc astropy}\footnote{www.astropy.org}, a community-developed core Python package for Astronomy {\sc astropy}, and  ``Aladin sky atlas'' developed at CDS, Strasbourg Observatory, France. We have also used the {\sc simbad} database and the {\sc vizier} catalogue access tool operated at CDS, Strasbourg (France), as well as NASA's Astrophysics Data System Bibliographic Services. This research used the facilities of the Canadian Astronomy Data Centre operated by the National Research Council of Canada with the support of the Canadian Space Agency. This research was made possible through the use of the AAVSO Photometric All-Sky Survey (APASS), funded by the Robert Martin Ayers Sciences Fund. Based on observations made with the NASA/ESA Hubble Space Telescope, and obtained from the Hubble Legacy Archive, which is a collaboration between the Space Telescope Science Institute (STScI/NASA), the Space Telescope European Coordinating Facility (ST-ECF/ESA) and the Canadian Astronomy Data Centre (CADC/NRC/CSA).




\bibliographystyle{mnras}
\bibliography{main_library} 

\bsp	
\label{lastpage}
\end{document}